\documentclass[]{aastex631}

\hypersetup{hypertexnames=false}
\shorttitle{Outflows of J0242+0049}
\shortauthors{Byun et al.}
\graphicspath{{./}{figures/}}

\begin{document}

\title{The Farthest Quasar Mini-BAL Outflow from its Central Source: VLT/UVES Observation of SDSS J0242+0049}

\author[0000-0002-3687-6552]{Doyee Byun}
\affiliation{Department of Physics, Virginia Tech, Blacksburg, VA 24061, USA}

\author[0000-0003-2991-4618]{Nahum Arav}
\affiliation{Department of Physics, Virginia Tech, Blacksburg, VA 24061, USA}

\author{Patrick B. Hall}
\affiliation{Department of Physics and Astronomy, York University, Toronto, ON
M3J 1P3, Canada}



\begin{abstract}

We analyze VLT/UVES observations of the quasar SDSS J024221.87+004912.6. We identify four absorption outflow systems: a \ion{C}{4} BAL at $v\approx -18,000\text{ km s}^{-1}$, and three narrower low-ionization systems with centroid velocities ranging from --1200 to --3500 $\text{ km s}^{-1}$. These outflows show similar physical attributes to the [\ion{O}{3}] outflows studied by \citet{2013MNRAS.436.2576L}. We find that two of the systems are energetic enough to contribute to AGN feedback, with one system reaching above $5\%$ of the quasar's Eddington luminosity. We also find that this system is at a distance of 67 kpc away from the quasar, the farthest detected mini-BAL absorption outflow from its central source to date. In addition, we examine the time variability of the BAL, and find that its velocity monotonically increases, while the trough itself becomes shallower over time.

\end{abstract}



\section{Introduction} \label{sec:intro}
Quasar absorption outflows are seen in a large fraction of quasar spectra ($\lesssim40\%$), often detected via blueshifted absorption troughs in the rest frame of quasars \citep{2003AJ....125.1784H,2008ApJ...672..108D,2008MNRAS.386.1426K}. These outflows are often mentioned as likely candidates for producing AGN feedback \citep[e.g.,][]{1998A&A...331L...1S,2004ApJ...608...62S,2009ApJ...699...89C,2018ApJ...857..121Y,2021arXiv210608337V}. According to theoretical models, outflow systems require a kinetic luminosity ($\dot{E}_k$) of at least $\sim0.5\%$ \citep{2010MNRAS.401....7H} or $\sim5\%$ \citep{2004ApJ...608...62S} of the quasar's Eddington luminosity ($L_{Edd}$) to contribute to AGN feedback. Outflow systems that fit these criteria have been found \citep[e.g.,][]{2009ApJ...706..525M,2013MNRAS.436.3286A,2020ApJS..247...37A,2015MNRAS.450.1085C,2019ApJ...876..105X,2020ApJS..247...38X,2020ApJS..247...42X,2020ApJS..247...39M,2020ApJS..249...15M}. \par
The kinetic luminosity of a quasar's outflow system is dependent on its distance from its central source ($R$), which we can find by measuring both the electron number density ($n_e$) and ionization parameter ($U_H$) \citep{2012ApJ...758...69B}. Our group and others have used this method to find the distances of outflow systems in the past \citep{2001ApJ...548..609D,2002ApJ...567...58D,2001ApJ...550..142H,2005ApJ...631..741G,2012ApJ...758...69B,2018ApJ...858...39X,2020ApJS..247...37A,2020ApJS..247...39M}. Using the ratios between excited and resonance state column densities of ionized species ($N_{ion}$) can lead us to a value of $n_e$ \citep{2018ApJ...857...60A}. This paper presents one such determination of the $R$ and $\dot{E}_k$ values of three outflow components found in the VLT/UVES spectrum of SDSS J024221.87+004912.6 (hereafter J0242+0049).\par
The analysis of J0242+0049 shown in this paper is based on data from the VLT/UVES Spectral Quasar Absorption Database (SQUAD) published by \cite{Murphy2019}, containing the spectra of 475 quasars. Analysis of more SQUAD objects will be conducted in the future.\par
The UVES data of J0242+0049 is from the program 075.B-0190(A) which \citet{2007ApJ...665..174H} used to identify a high velocity \ion{C}{4} broad absorption line (BAL) at $z\approx 1.88$ $(v\approx-18,000 \text{ km s}^{-1})$, as well as two mini-BAL outflows and one narrow absorption line (NAL) outflow at lower velocities; all four of which we have identified independently. Comparing the UVES spectrum to SDSS spectra from previous epochs, \citet{2007ApJ...665..174H} have identified a shift in the velocity of the high velocity BAL, which could potentially be explained by acceleration. They have also found potential line locking in the \ion{Si}{4} absorption doublets of the two lower velocity mini-BAL systems. In addition to the analysis of the UVES data, we conduct a follow up to their observation of the velocity shift using SDSS observation data from more recent epochs.\par
This paper is structured as follows. Section \ref{sec:data} discusses the observation of J0242+0049, as well as the data acquisition process. In Section \ref{sec:analysis}, we present the ionic column density measurements, and the process of finding $n_e$ and $U_H$. Section \ref{sec:results} shows the results of the analysis, including the energetics parameters of the outflow systems. We also show observations of the high velocity BAL from recent SDSS epochs. Section \ref{sec:discussion} provides a discussion of the results, and Section \ref{sec:conclusion} summarizes and concludes the paper. For this analysis, we adopt a cosmology of $h=0.696$, $\Omega_m = 0.286$, and $\Omega_\Lambda = 0.714$ \citep{Bennett_2014}, and use the Python astronomy package Astropy \citep{astropy:2013,astropy:2018} for cosmological calculations.

\section{Observation, Data Acquisition, and Line Identification} \label{sec:data}
\begin{figure*}
    \centering
    \plotone{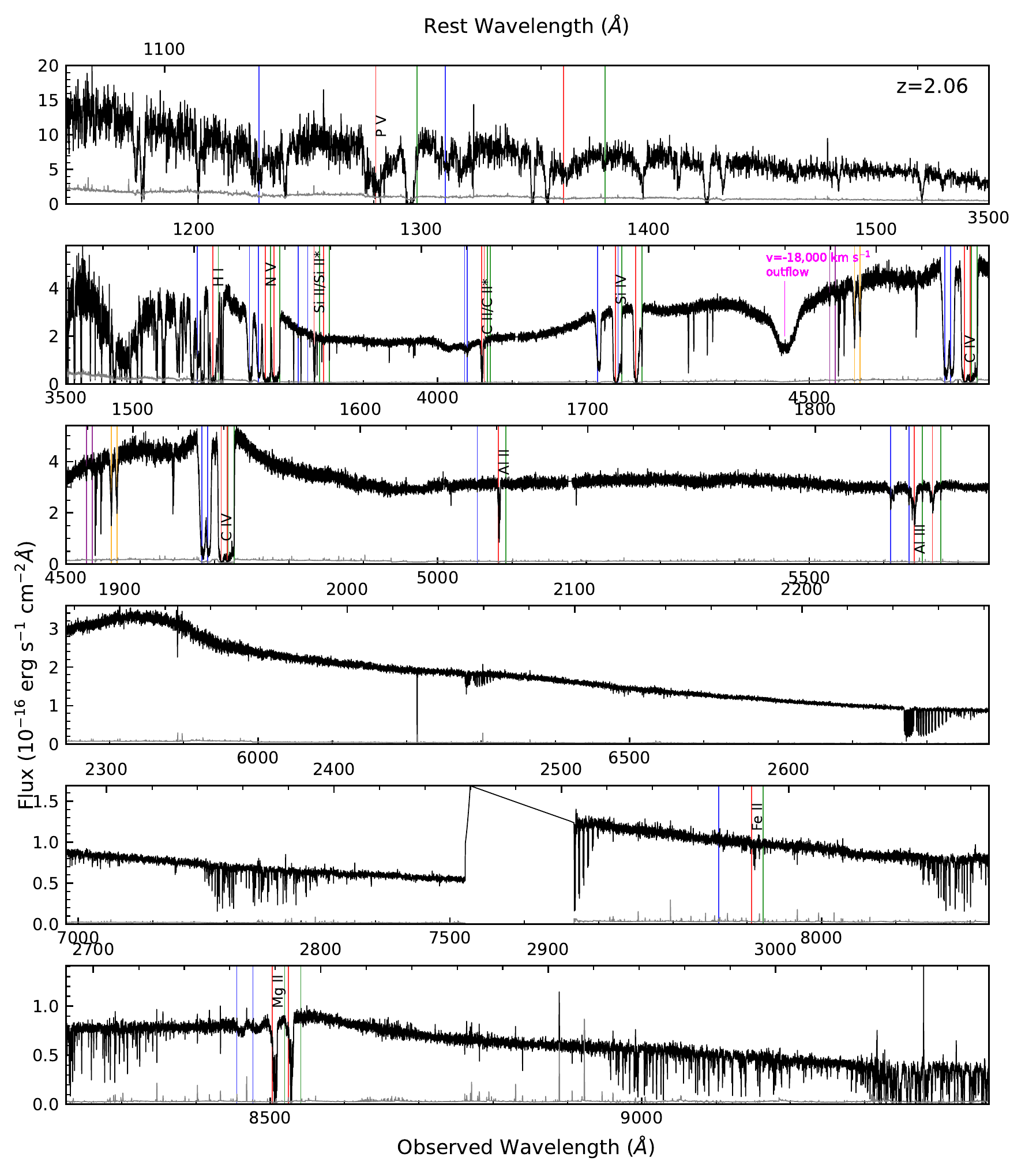}
    \caption{Normalized flux of J0242+0049 multiplied by the emission model by \citet{Murphy2019}, based on the SQUAD data set. The flux has been scaled to match the BOSS spectrum from the epoch of MJD=57758 (Jan. 5, 2017) at observed wavelength $\lambda = 6500 \AA$. The black curve represents the flux, and the gray shows the error in flux. The green, red, and blue vertical lines mark absorption troughs of outflow systems S1, S2, and S3, respectively, while the S4 \ion{C}{4} BAL is labeled in magenta. Systems A and B are marked in orange and purple respectively. Note that the absorption troughs for S1 are significantly narrower when compared to those of S2 and S3.}
    \label{fig:fluxplot}
\end{figure*}

The quasar J0242+0049 (J2000: RA=02:42:22, DEC=+00:49:12.6; z=2.06) \citep{2018A&A...613A..51P} was observed in September 5, 2005 with the VLT/UVES as part of the program 075.B-0190(A), with resolution $R\simeq 40,000$ and wavelength coverage from 3291 to 9300 $\AA$ \citep{2007ApJ...665..174H}. The systemic redshift z=2.06 given by \cite{Murphy2019} is consistent with the value we find based on the Mg II emission line in the SDSS spectrum of the MJD=57758 epoch. The spectral data was reduced and normalized by its continuum and emission by \cite{Murphy2019} as part of their SQUAD database. Broad and narrow absorption lines have been found in the spectrum of J0242+0049 by \citet{2007ApJ...665..174H}, which we identify here as NAL S1 at --1200 km s$^{-1}$ (Ly $\alpha$ FWHM = 240 km s$^{-1}$), mini-BAL S2 at -1800 km s$^{-1}$ (\ion{N}{5} FWHM = 900 km s$^{-1}$), mini-BAL S3 at --3500 km s$^{-1}$ (\ion{N}{5} FWHM = 720 km s$^{-1}$), and the aforementioned BAL S4 at --18,000 km s$^{-1}$, as shown in the full spectrum in Figure \ref{fig:fluxplot}. Following \citet{1991ApJ...373...23W}, a BAL is a continuous absorption feature below 0.9 normalized intensity over 2000 km $s^{-1}$, a mini-BAL is the same but between 500 and 2000 km s$^{-1}$ \citep{2004ASPC..311..203H}, and a NAL is an absorption feature with width below 500 km s$^{-1}$. We measure the width of S4 at 0.9 normalized intensity to be 2200 km s$^{-1}$, which is above the threshold of a BAL, with balnicity index as defined by \citet{1991ApJ...373...23W} of 660 km s$^{-1}$. \citet{2021ApJ...907...84C} have identified four \ion{C}{4} absorption systems, two of which coincide with systems S3 and S4. We label the other two as systems A and B, and show them in Figure \ref{fig:fluxplot}. The focus of this paper is on the four systems S1, S2, S3, and S4. We do not discuss systems A and B because they only show absorption in \ion{C}{4}, which does not lend itself to further analysis.\par
\begin{figure*}
\gridline{\fig{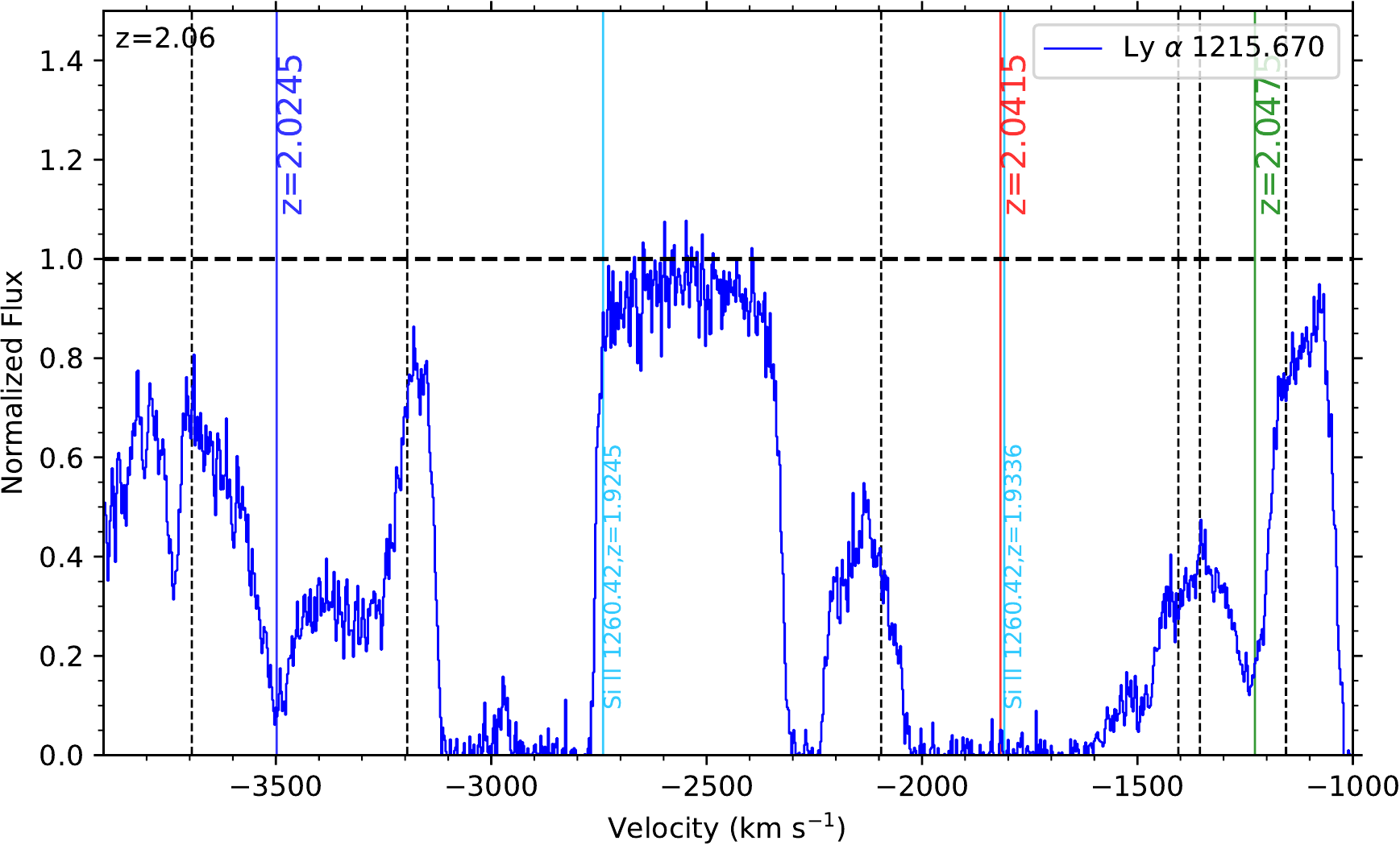}{0.33\textwidth}{(a) Ly $\alpha$}
          \fig{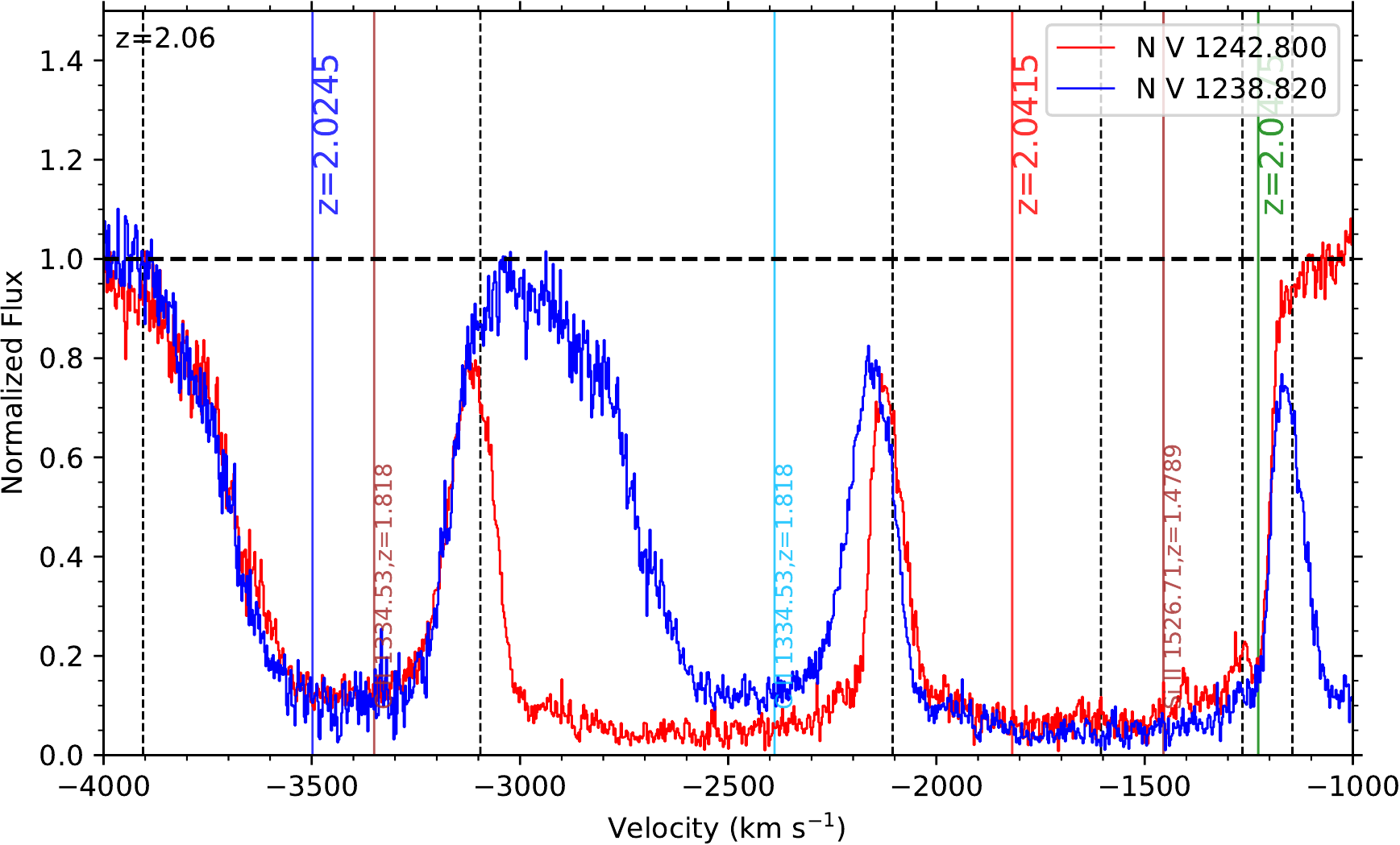}{0.33\textwidth}{(b) \ion{N}{5}}
          \fig{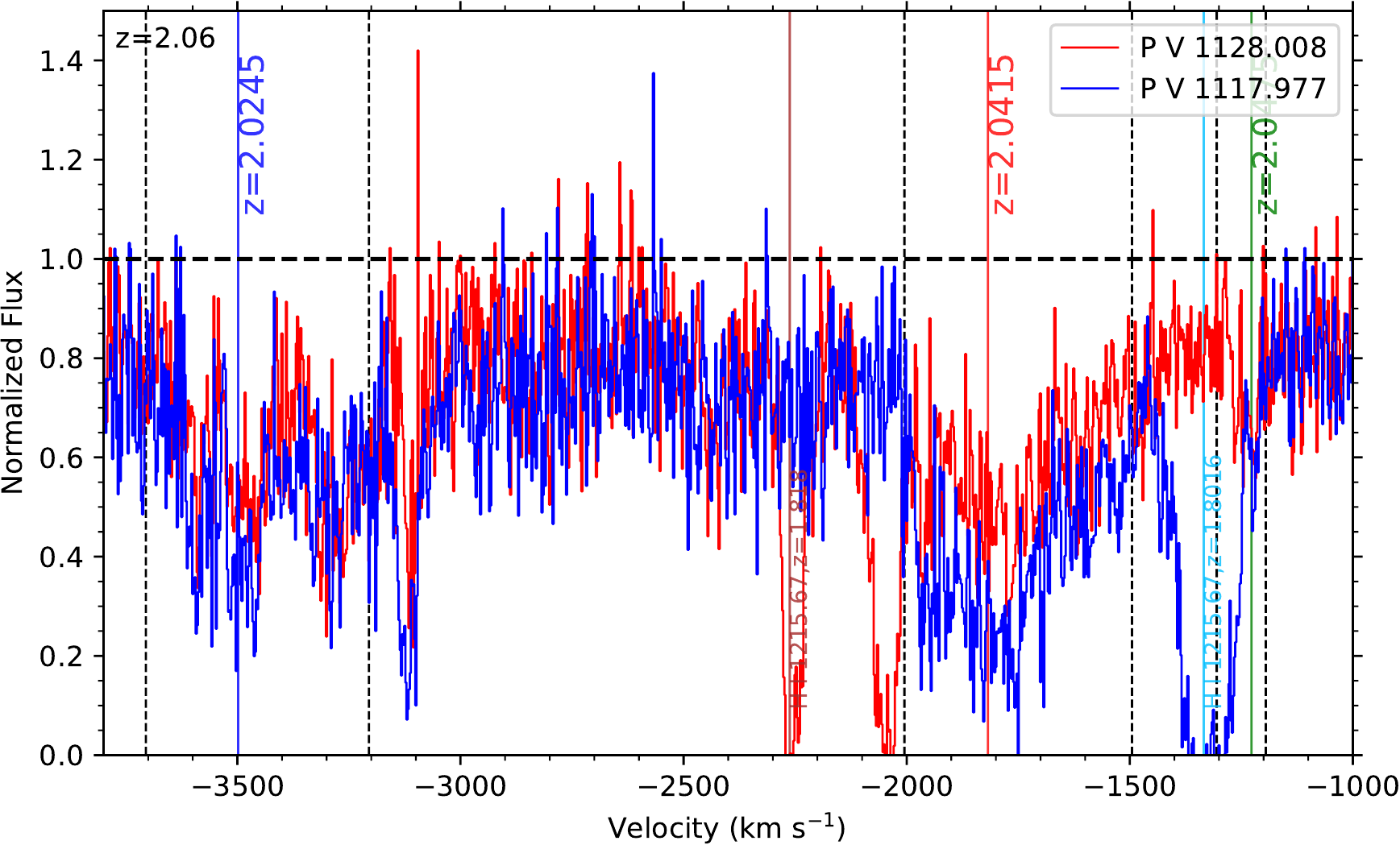}{0.33\textwidth}{(c) \ion{P}{5}}}
\gridline{\fig{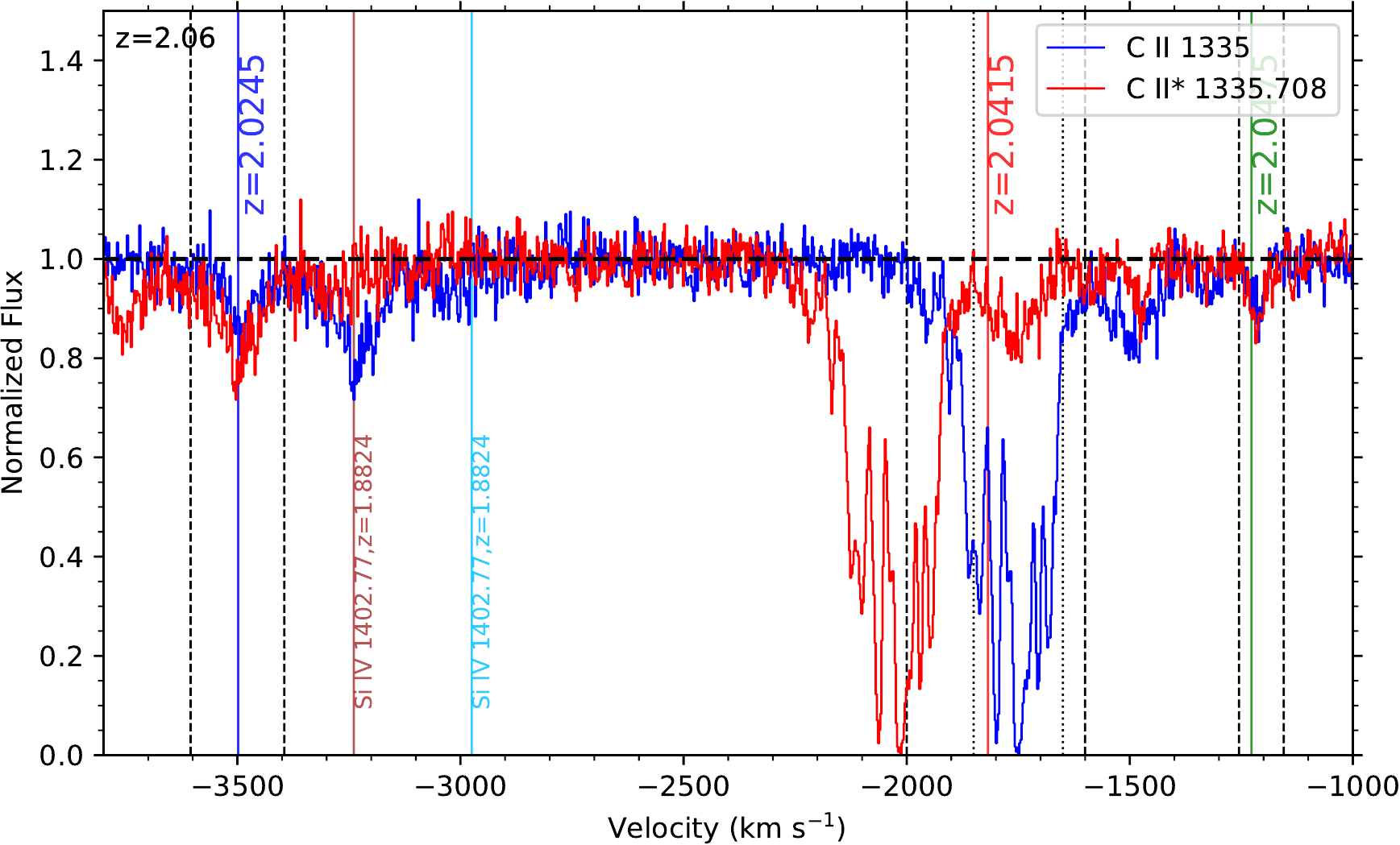}{0.33\textwidth}{(d) \ion{C}{2}}
          \fig{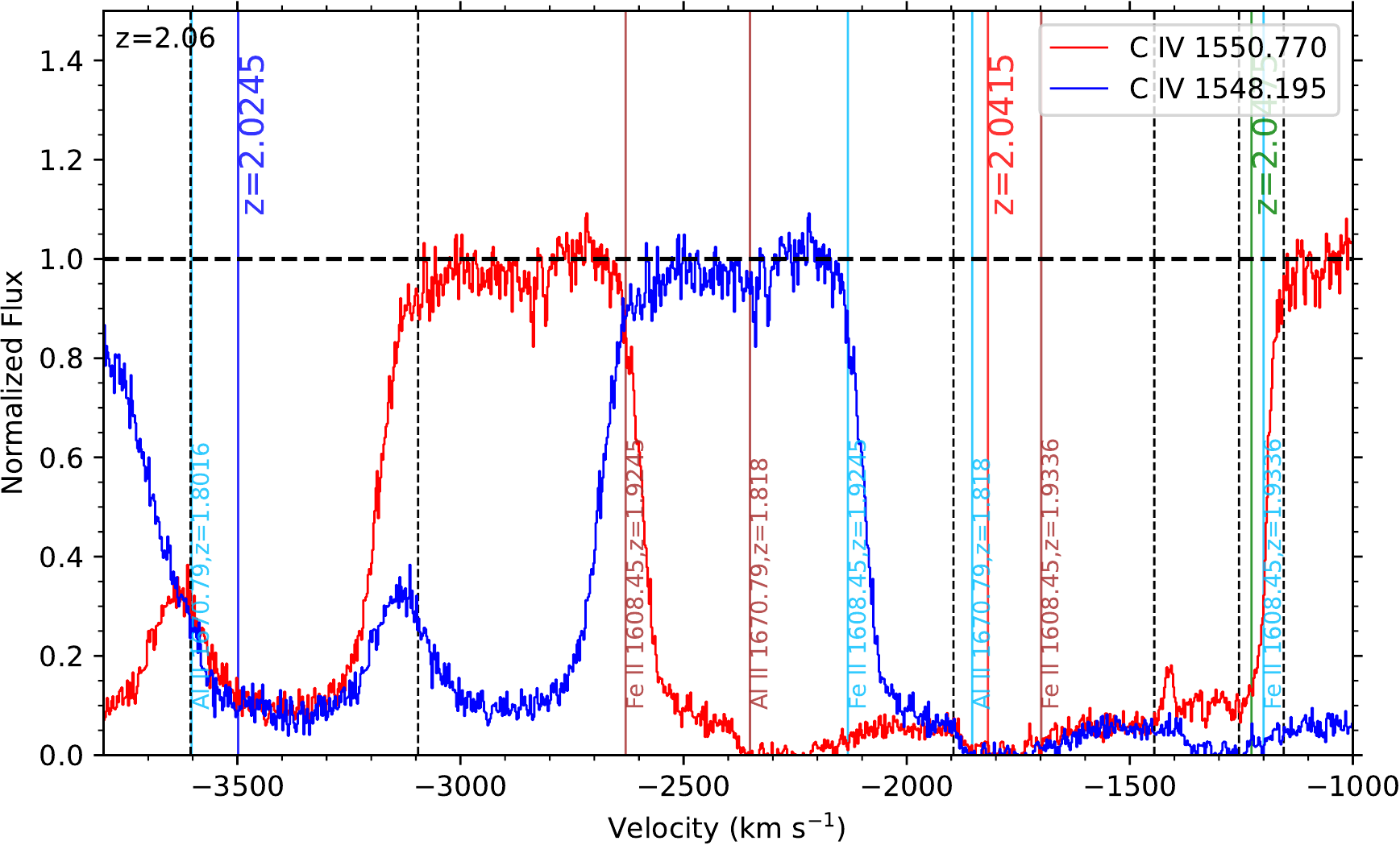}{0.33\textwidth}{(e) \ion{C}{4}}
          \fig{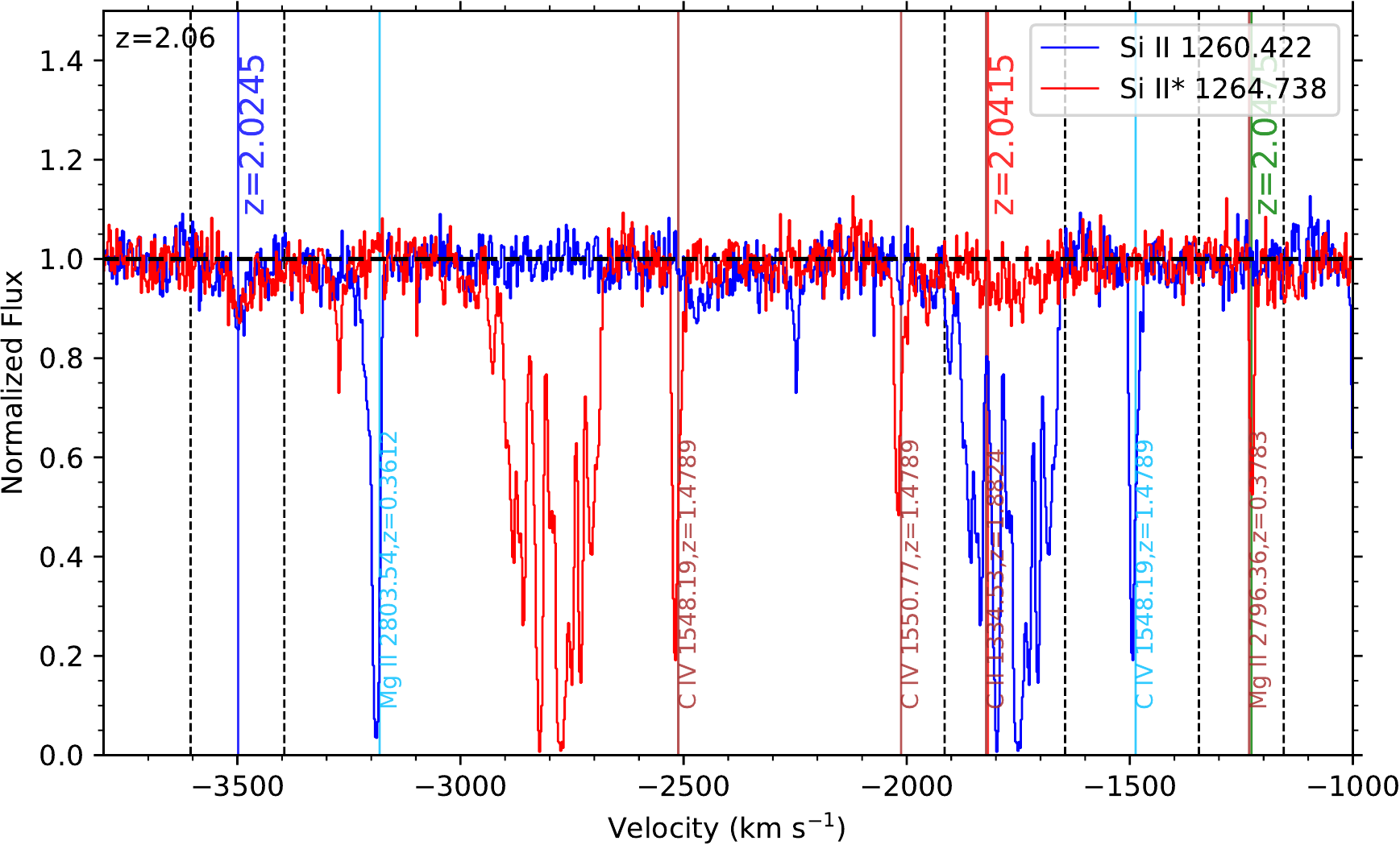}{0.33\textwidth}{(f) \ion{Si}{2}}}
\gridline{\fig{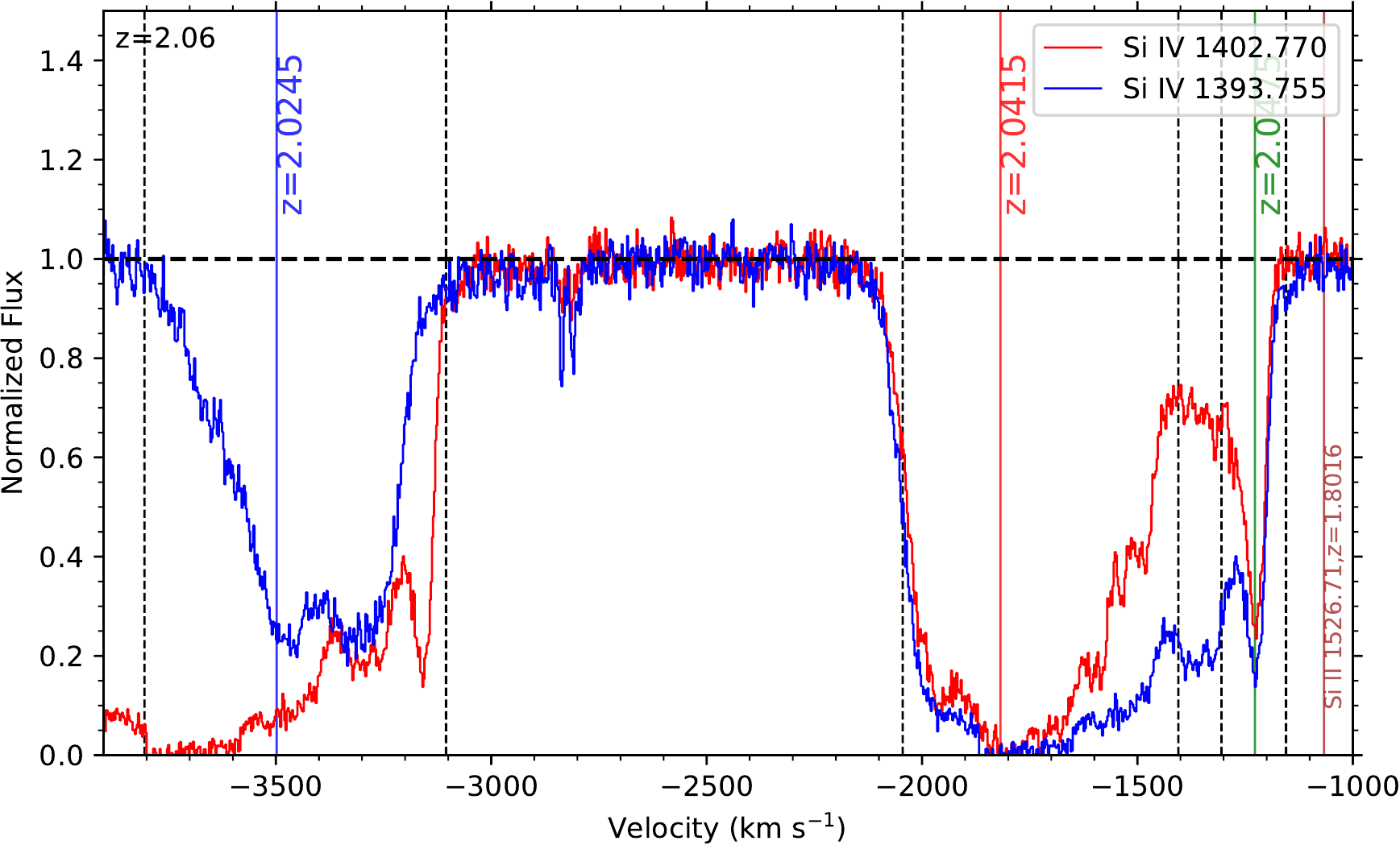}{0.33\textwidth}{(g) \ion{Si}{4}}
          \fig{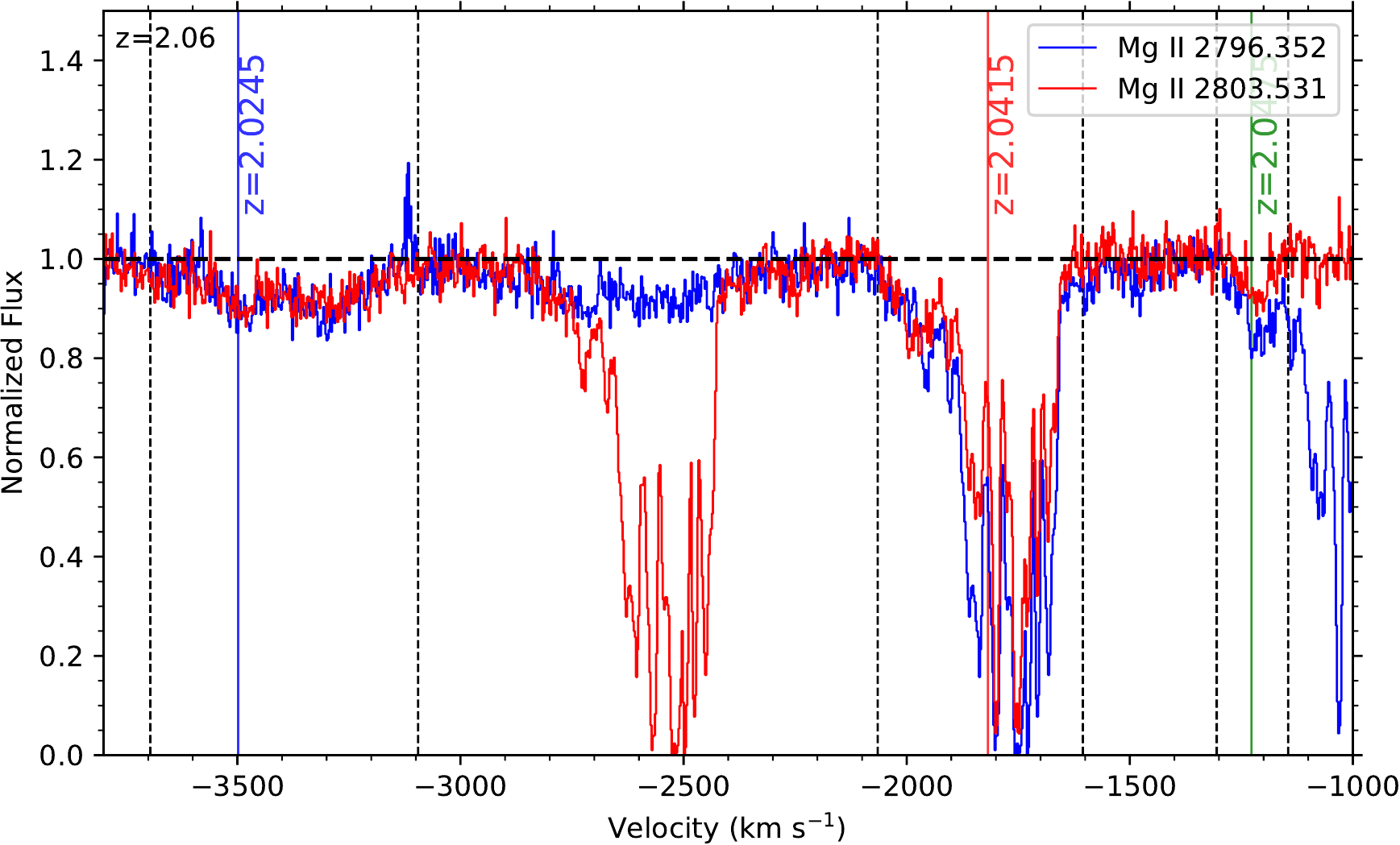}{0.33\textwidth}{(h) \ion{Mg}{2}}
          \fig{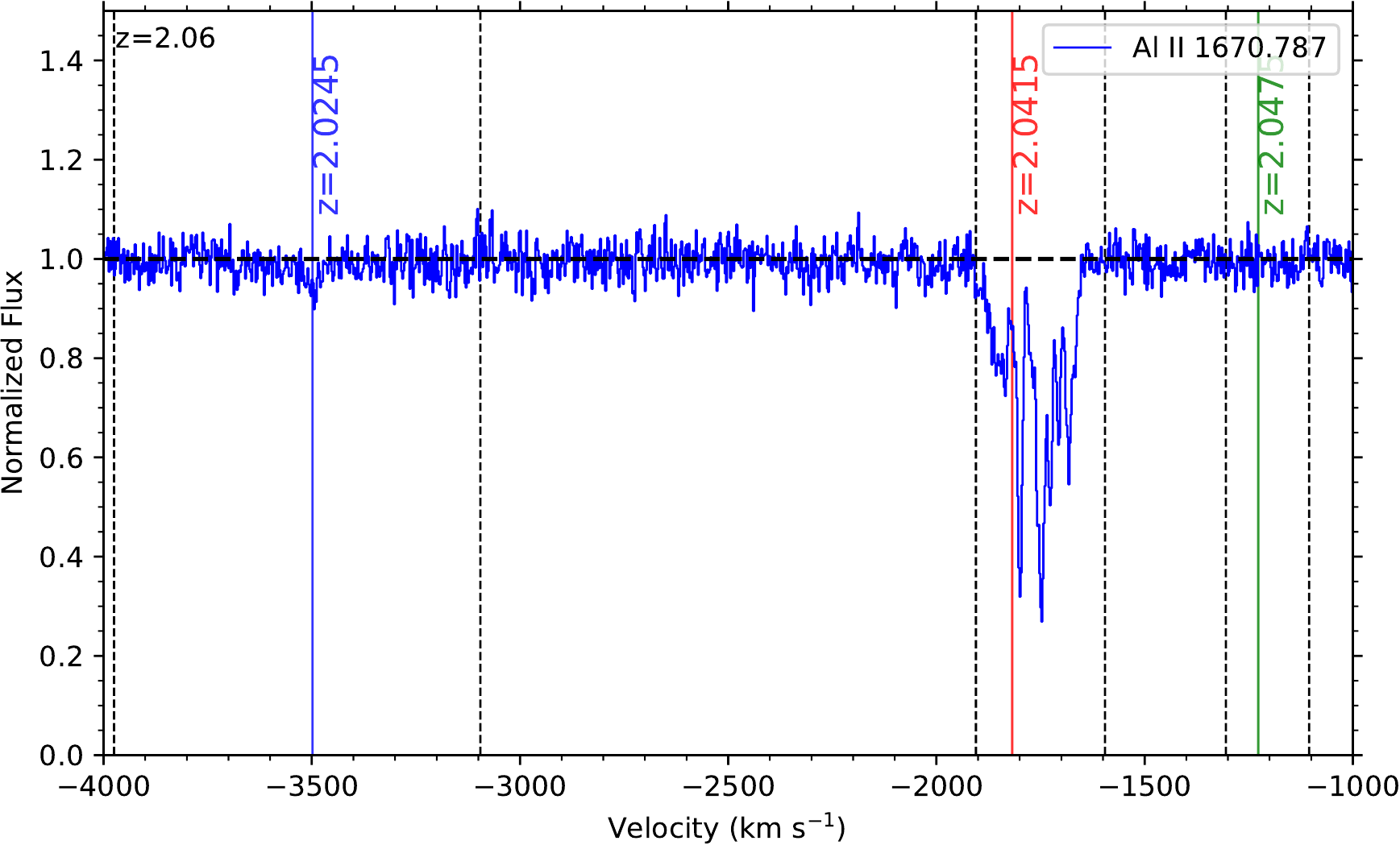}{0.33\textwidth}{(i) \ion{Al}{2}}}
\gridline{\fig{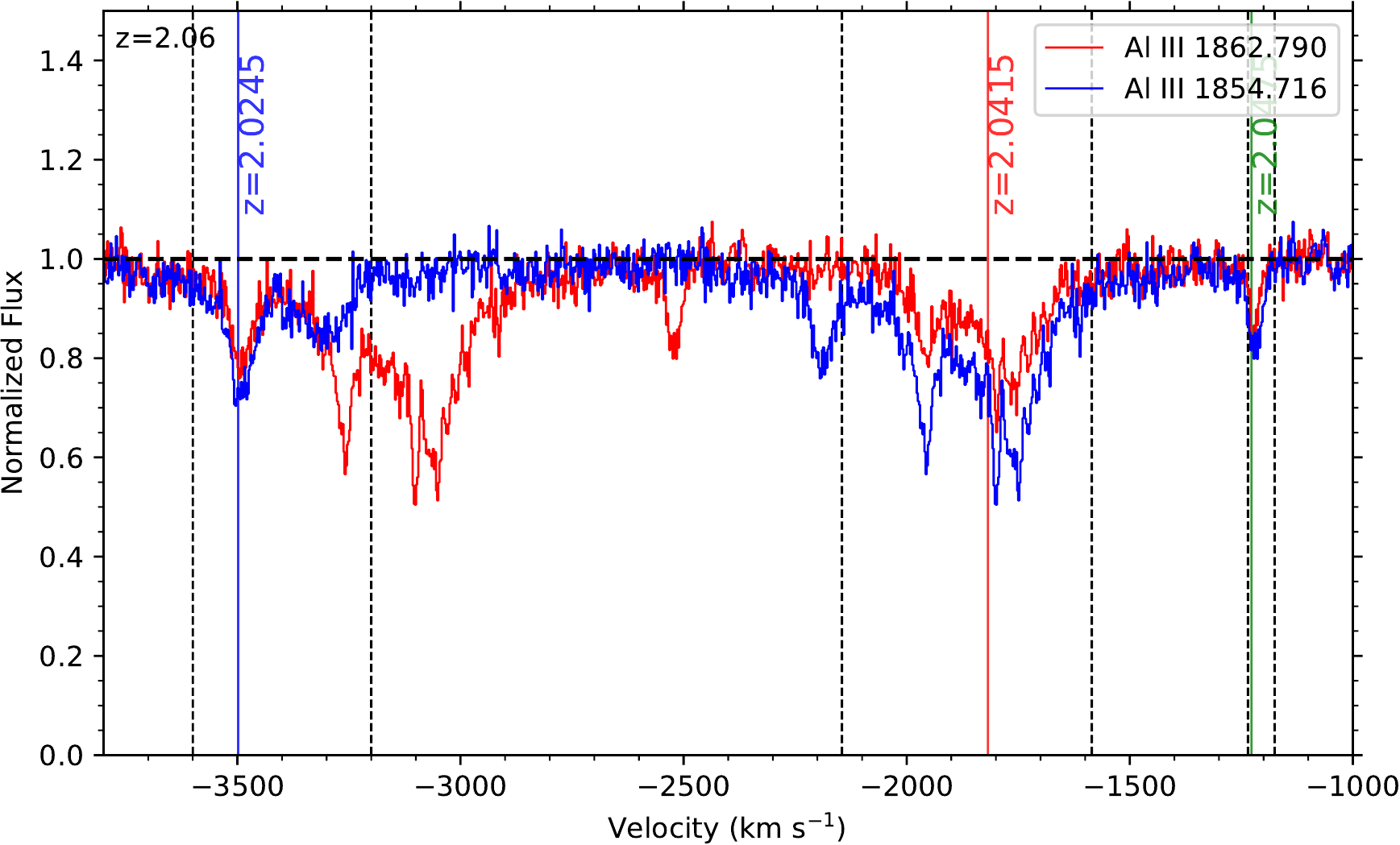}{0.33\textwidth}{(j) \ion{Al}{3}}
          \fig{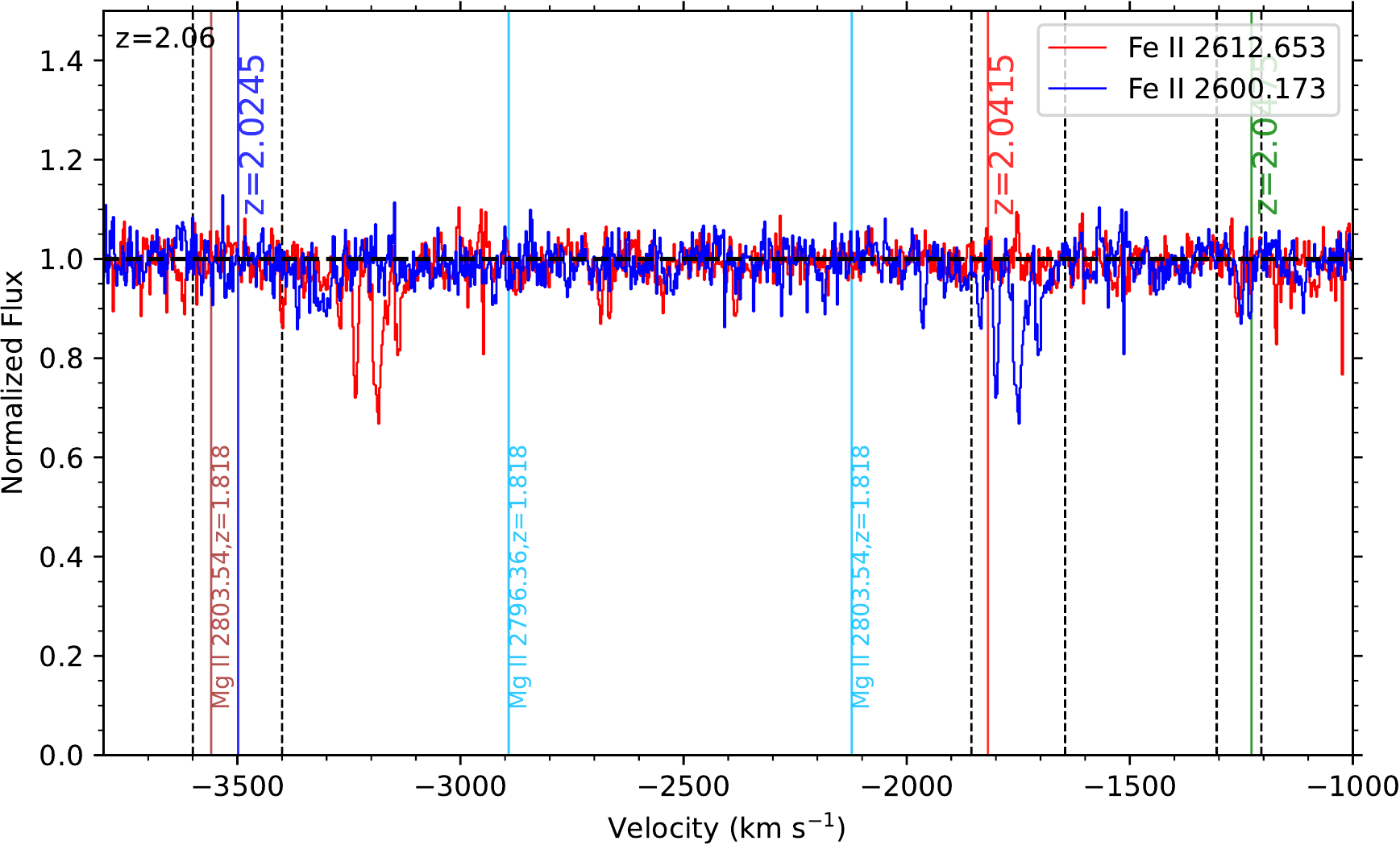}{0.33\textwidth}{(k) \ion{Fe}{2}}
          \fig{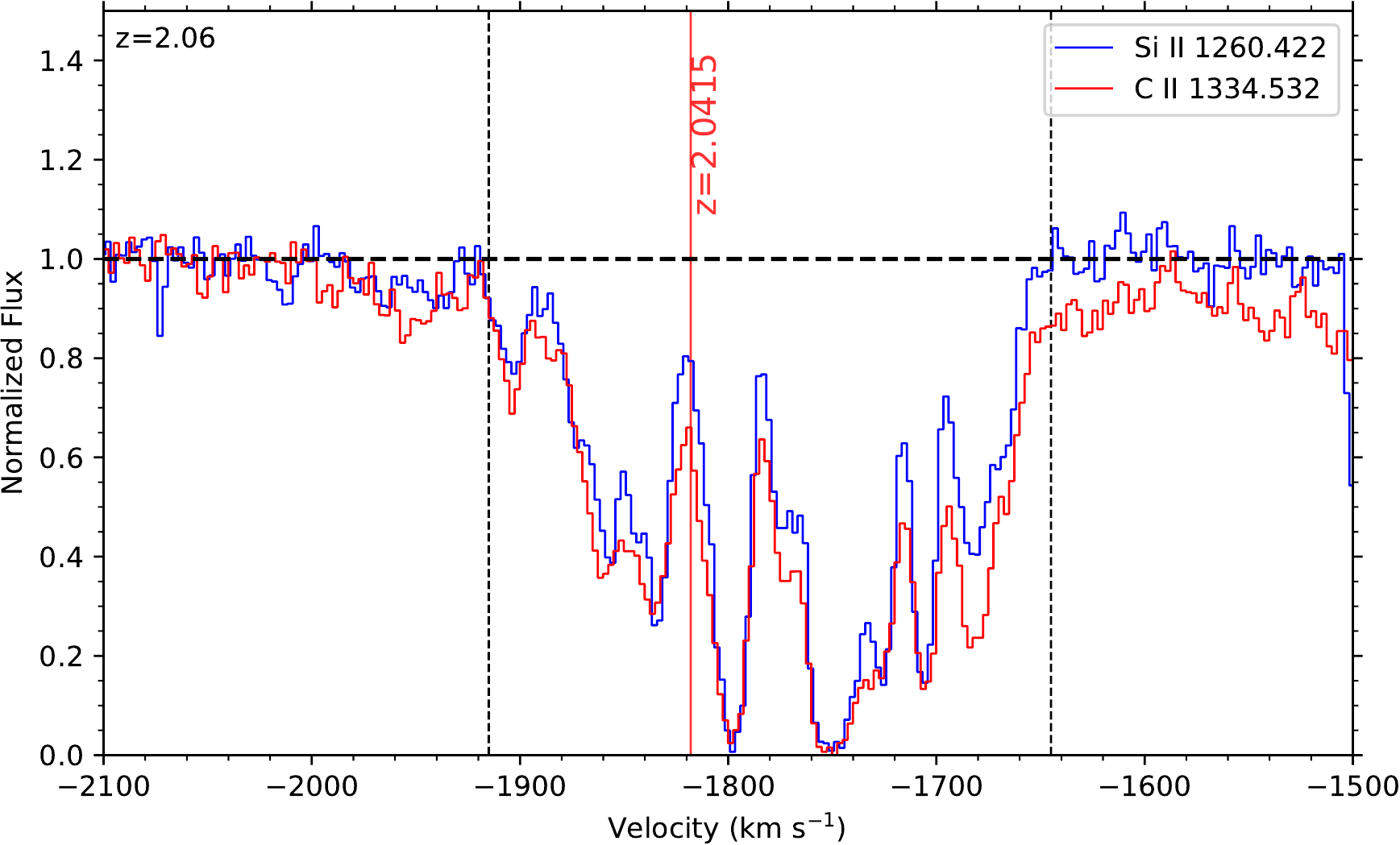}{0.33\textwidth}{(l) \ion{C}{2} \& \ion{Si}{2}}}
\caption{Normalized spectrum plotted in velocity space for each ion in the absorption systems. The green, red, and blue vertical lines represent the velocity of systems S1, S2, and S3, respectively. The dotted vertical lines show the integration ranges used for the calculation of the ionic column densities. The horizontal dashed line represents the continuum level. Intervening absorption systems that contaminate the blue spectra are marked with cyan vertical lines, while intervening systems contaminating the red spectra are marked with brown vertical lines. Note in Plot (f) that the S1 integration range for \ion{Si}{2} 1265\AA\space is contaminated with \ion{Mg}{2} 2796\AA\space absorption of the z=0.3783 intervening system. Plot (l) shows the structure of the S2 absorption trough of \ion{Si}{2} and \ion{C}{2}, on a narrower velocity scale.}
\label{fig:vcut}
\end{figure*}

The outflows show absorption from low ionization species such as \ion{Si}{2}, \ion{C}{2}, and \ion{Fe}{2}, as well as lines of Ly $\alpha$, \ion{C}{4}, \ion{N}{5}, \ion{P}{5}, \ion{Mg}{2}, \ion{Al}{2}, and \ion{Al}{3}. For the purpose of measuring the ionic column densities, we convert the normalized spectrum data from wavelength to velocity space via the systemic redshift of the quasar, as shown in Figure \ref{fig:vcut}. Note that S2 appears to be composed of at least seven sub-components, as seen in Plot (l) of Figure \ref{fig:vcut}. The components are blended in the absorption troughs of \ion{C}{4} and \ion{Si}{4}, and due to the shallowness of the \ion{C}{2}* troughs, it is impossible to decompose it into the different sub-components. For this reason, they are treated as a singular absorption system for the sake of the analysis in this paper.\par
For the velocity shift analysis, SDSS spectra from MJD=52177, 52199, 55455, and 57758 were retrieved and corrected for galactic extinction with $E(B-V)=0.0269$ \citep{2011ApJ...737..103S}. The spectra from both the BOSS and SDSS spectographs have spectral resolutions of $R\approx2000$ \citep{2010AJ....139.2360S,2013AJ....146...32S,2018A&A...613A..51P}. More details on the SDSS spectra can be found in Table \ref{table:sdss_spectra}.\par
\begin{deluxetable}{lccccc}
\tabletypesize{\scriptsize}
\tablenum{1}
\tablecaption{SDSS Spectra Information\label{table:sdss_spectra}}
\tablehead{\colhead{Epoch in MJD}&\colhead{Spectrograph}&\colhead{Plate}&\colhead{Fiber}&\colhead{Observed Date}&\colhead{Wavelength Coverage ($\AA$)}}
\startdata
\text{52177}&SDSS&707&332&Sep. 25, 2001&3824--9215\\
\text{52199}&SDSS&706&617&Oct. 17, 2001&3820--9202\\
\text{55455}&BOSS&4240&754&Sep. 16, 2010&3590--10382\\
\text{57758}&BOSS&9381&79&Jan. 5, 2017&3573--10334
\enddata
\end{deluxetable}
\section{Analysis} \label{sec:analysis}
\subsection{Ionic Column Density} \label{subsec:coldensity}
To find the physical characteristics of the outflow systems, we first find the column densities of the observed ions ($N_{ion}$). The simplest method of measuring column densities is by assuming the apparent optical depth (AOD) of a uniformly covered homogeneous source, as demonstrated by \cite{Savage1991}. When calculating column density under this assumption, we first assume the relation between intensity and optical depth as follows \citep[see equation 1 of][]{Savage1991}:\\
\begin{equation}
    I(\lambda)=I_0(\lambda)e^{-\tau(\lambda)} 
\end{equation}
where $I(\lambda)$ is the intensity, $I_0(\lambda)$ is the intensity without absorption, and $\tau(\lambda)$ is the optical depth as a function of wavelength. When writing optical depth as a function of outflow velocity, it has a relation with column density $N(v)$ of \citep[see equation 8 of][]{Savage1991}:\\
\begin{equation}
    \tau(v)=\frac{\pi e^2}{m_e c}f\lambda N(v)
\end{equation}
where $m_e$ is the mass of an electron, $e$ is the elementary charge, and $f$ and $\lambda$ are the oscillator strength and wavelength of the transition line, respectively. Finding $N(v)$ and integrating it over the velocity range of the absorption trough yields the column density based on the AOD assumption. The AOD method is used to find lower limits of $N_{ion}$ for singlets or contaminated doublets, or upper limits when there are no discernible absorption troughs.\par
When there are multiple lines of the same ion and energy state, we can use the partial covering (PC) method, which assumes a homogeneous source partially covered by the outflow \citep{1997ASPC..128...13B,1999ApJ...524..566A,Arav1999}, and solves for a velocity dependent covering factor \citep{DeKool2002,2005ApJ...620..665A}, to improve our measurements by taking phenomena such as non-black saturation into account \citep{2011ApJ...739....7E,Borguet2012}. When calculating the PC based column density of an ion with a doublet of transition lines, we find the covering fraction $C(v)$ via the following relations \citep[see equations 2 \& 3 of][]{2005ApJ...620..665A}
\begin{eqnarray}
    I_R(v)-[1-C(v)]=C(v)e^{-\tau(v)}\\
    I_B(v)-[1-C(v)]=C(v)e^{-2\tau(v)}
\end{eqnarray}
where $I_R(v)$ and $I_B(v)$ are the normalized intensities of the red and blue absorption features respectively, and $\tau(v)$ is the optical depth of the red component.\par
We choose integration ranges that cover visible absorption in the data, as can be seen in Figure 2, while minimizing the effects of blending and contamination. For instance, for Si IV, we use the blue line for S3 and the red line for S2. Si II* of S1 shows contamination due to an intervening absorption feature, so we use the measured column density as an upper limit for the sake of our analysis. C IV of S2 is heavily blended between the red and blue features, so we choose a velocity range in which the blue and red spectra do not overlap with each other in order to find a lower limit of the column density.\par
Attempting a Gaussian fit of the C IV absorption of S2 yields a poor fit due to the saturation of the trough. Calculating the column density based on the fit results in a lower limit of $2400\times10^{12}\text{cm}^{-2}$, compared to the measured lower limit of $3900\times10^{12}\text{cm}^{-2}$. This difference does not affect the solution of the hydrogen column density and photoionization parameter as described in Section \ref{subsec:photoionization}.\par
The measured column density values can be found in Table \ref{table:coldensity}. Note that most adopted values in Table \ref{table:coldensity} are upper or lower limits. The errors in the column densities are propagated from the errors in the normalized flux from the data, binned along with the data into segments of $\Delta v=10$ km s$^{-1}$ for numerical integration. 20\% error is added in quadrature for the column density values adopted for photoionization analysis (see last column of Table \ref{table:coldensity}) to take into account the uncertainty in the modeled continuum level \citep{2018ApJ...858...39X}.\par
\begin{deluxetable}{lccc}
\tablenum{2}
\tablecaption{J0242+0049 Outflow Ionic Column Densities\label{table:coldensity}}
\tabletypesize{\scriptsize}
\tablewidth{\columnwidth}
\tablehead{\colhead{Troughs}&\colhead{AOD}&\colhead{PC}&\colhead{Adopted}} 

\startdata
\multicolumn{4}{c}{S1, $v=-1200\text{ km s}^{-1}$}\\
\hline
\text{H I} &$177.0_{-1.8}^{+1.9}$  &&$>180_{-40}$\\
\text{\ion{N}{5}} &$467_{-5}^{+5}$&&$>470_{-90}$\\
\text{\ion{P}{5}} &$56_{-4}^{+4}$&&$>50_{-10}$\\
\text{\ion{C}{2} total}&$26_{-2}^{+2}$&&$>26_{-5}$\\
\text{\ion{C}{2} 1335} &$10.8_{-1.2}^{+1.5}$  &&\\
\text{\ion{C}{2}* 1336} &$14.7_{-1.3}^{+1.5}$  &&\\
\text{\ion{C}{4}} &$350_{-4}^{+5}$  &&$>350_{-70}$\\
\text{\ion{Si}{2} total} & $5.5_{-0.4}^{+0.3}$&&$<5.5^{+1.2}$\\
\text{\ion{Si}{2} 1260} & $1.3_{-0.2}^{+0.3}$&&$<1.3^{+0.4}$\\
\text{\ion{Si}{2}* 1265} & $2.8_{-0.2}^{+0.2}$&&$<2.8^{+0.6}$\\
\text{\ion{Si}{4}} & $92.6_{-1.0}^{+1.0}$&&$>90_{-20}$\\
\text{\ion{Mg}{2}} & $2.7_{-0.3}^{+0.3}$&&$>2.7_{-0.6}$\\
\text{\ion{Al}{2}} & $0.3_{-0.08}^{+0.09}$&&$<0.3^{+0.1}$\\
\text{\ion{Al}{3}} & $3.3_{-0.3}^{+0.3}$&$4.9_{-0.5}^{+0.8}$&$4.9_{-1.1}^{+1.3}$\\
\text{\ion{Fe}{2}} & $1.9_{-0.3}^{+0.4}$&&$<1.9^{+0.5}$\\
\hline
\multicolumn{4}{c}{S2, $v=-1800\text{ km s}^{-1}$}\\
\hline
\text{\ion{H}{1}} &$1680_{-10}^{+180}$  &&$>1680_{-340}$\\
\text{\ion{N}{5}} &$4620_{-30}^{+30}$&&$>4620_{-920}$\\
\text{\ion{P}{5}} &$450_{-10}^{+10}$&&$>450_{-90}$\\
\text{\ion{C}{2} total} &$740_{-10}^{+90}$  &&$>740_{-150}$\\
\text{\ion{C}{2} 1335} &$690_{-10}^{+90}$  &&\\
\text{\ion{C}{2}* 1336} &$50_{-2}^{+2}$  &&\\
\text{\ion{C}{4}} &$3910_{-20}^{+330}$  &&$>3910_{-780}$\\
\text{\ion{Si}{2} total} & &&$>80_{-20}$\\
\text{\ion{Si}{2} 1260} & $77.6_{-1.1}^{+1.3}$&&$>80_{-20}$\\
\text{\ion{Si}{2}* 1265} & $2.9_{-0.3}^{+0.3}$&&$<3^{+0.7}$\\
\text{\ion{Si}{4}} & $1410_{-10}^{+100}$&&$>1410_{-280}$\\
\text{\ion{Mg}{2}} & $84_{-0.9}^{+1.0}$&$90.7_{-1.0}^{+1.0}$&$90_{-20}^{+20}$\\
\text{\ion{Al}{2}} & $10.3_{-0.1}^{+0.2}$&&$>10_{-2}$\\
\text{\ion{Al}{3}} & $48.4_{-0.9}^{+0.9}$&$55.6_{-0.8}^{+0.9}$&$55_{-10}^{+10}$\\
\text{\ion{Fe}{2} total} &&&$>12_{-2.5}$\\
\text{\ion{Fe}{2} 2600} & $12.2_{-0.5}^{+0.5}$&&$>12_{-2.5}$\\
\text{\ion{Fe}{2}* 2612} & $1.0_{-0.5}^{+0.5}$&&$<1.0^{+0.5}$\\
\hline
\multicolumn{4}{c}{S3, $v=-3500\text{ km s}^{-1}$}\\
\hline
\text{\ion{H}{1}} &$417.7_{-3.4}^{+3.7}$  &&$>420_{-80}$\\
\text{\ion{N}{5}} &$3780_{-20}^{+20}$&&$>3780_{-760}$\\
\text{\ion{P}{5}} &$390_{-10}^{+10}$&&$>390_{-80}$\\
\text{\ion{C}{2} total} &$94.2_{-3.3}^{+3.5}$  &&$>90_{-20}$\\
\text{\ion{C}{2} 1335} &$32_{-2.2}^{+2.4}$  &&\\
\text{\ion{C}{2}* 1336} &$62.1_{-2.5}^{+2.5}$  &&\\
\text{\ion{C}{4}} &$2180_{-10}^{+10}$  &&$>2180_{-440}$\\
\text{\ion{Si}{2} total} & $5.1_{-0.3}^{+0.4}$&&$>5.1_{-1.1}$\\
\text{\ion{Si}{2} 1260} & $2.5_{-0.3}^{+0.3}$&&\\
\text{\ion{Si}{2}* 1265} & $2.6_{-0.2}^{+0.2}$&&\\
\text{\ion{Si}{4}} & $285.8_{-1.4}^{+1.5}$&&$>290_{-60}$\\
\text{\ion{Mg}{2}} & $17.9_{-0.6}^{+0.6}$&&$>18_{-4}$\\
\text{\ion{Al}{2}} & $2_{-0.2}^{+0.2}$&&$<2^{+0.4}$\\
\text{\ion{Al}{3}} & $19.7_{-0.4}^{+0.4}$&&$>20_{-4}$\\
\text{\ion{Fe}{2} total} & $6.7_{-0.9}^{+0.7}$&&$<6.7^{+1.5}$\\
\text{\ion{Fe}{2} 2600} & $2.2_{-0.5}^{+0.4}$&&\\
\text{\ion{Fe}{2}* 2612} & $4.5_{-0.7}^{+0.6}$&&
\enddata\tablecomments{Units are in $10^{12}\text{cm}^{-2}$.  Values have been calculated by numerical integration over bins with width $\Delta v = 10 \text{km s}^{-1}$. Note that most of the adopted values are upper or lower limits.}

\end{deluxetable}
\subsection{Photoionization Analysis} \label{subsec:photoionization}
We use a grid of photoionization models created using the spectral synthesis code Cloudy (version c17.00) \citep{2017RMxAA..53..385F}, in order to find the Hydrogen column density ($N_H$) and ionization parameter ($U_H$) that best fit the measured ionic column densities, following the method of previous works \citep[e.g.,][]{2019ApJ...876..105X,2018ApJ...865...90M,2020ApJS..247...39M}. \par
\begin{figure*}
    \centering
    \gridline{\fig{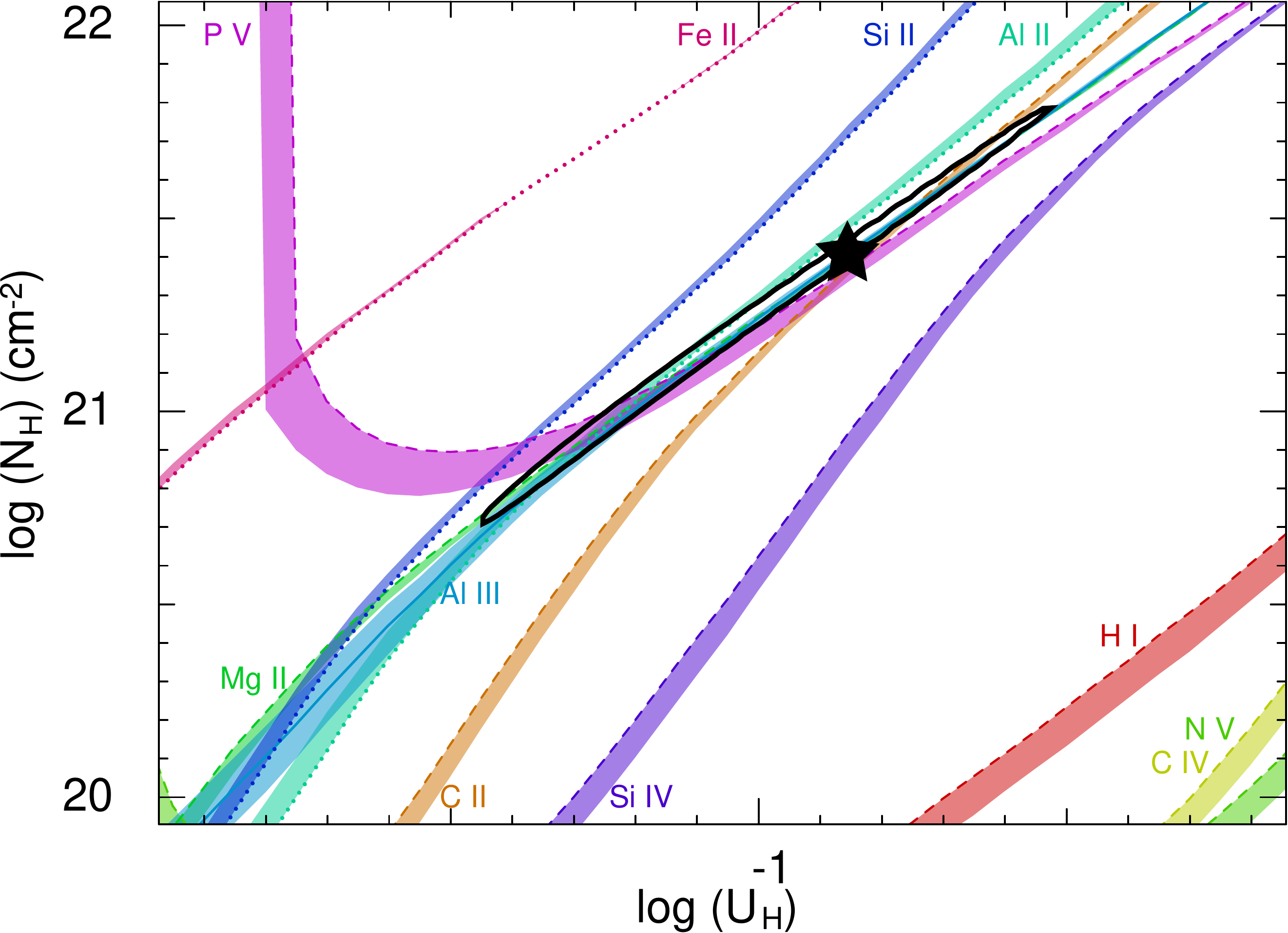}{0.33\textwidth}{(a)}
              \fig{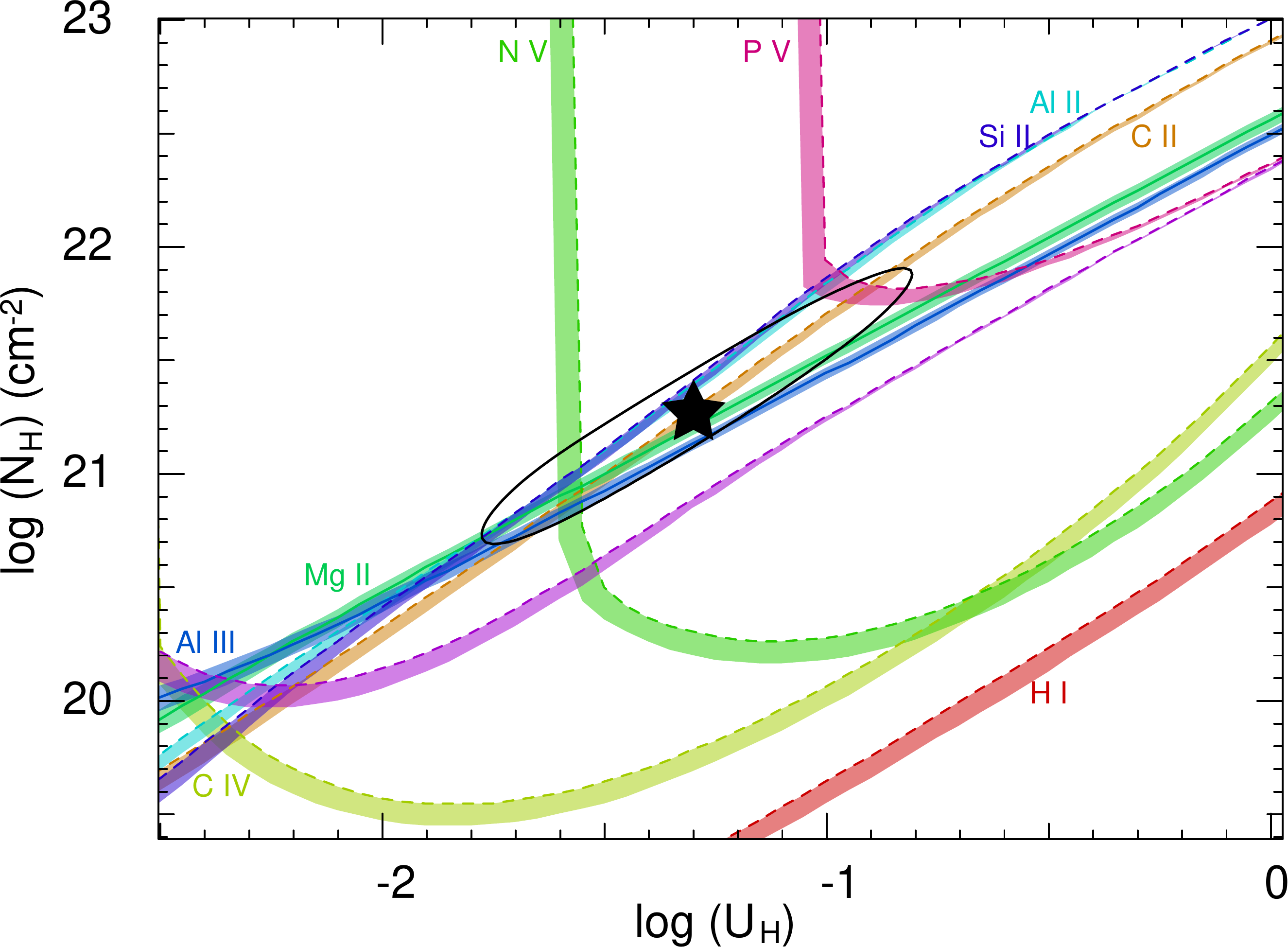}{0.33\textwidth}{(b)}
              \fig{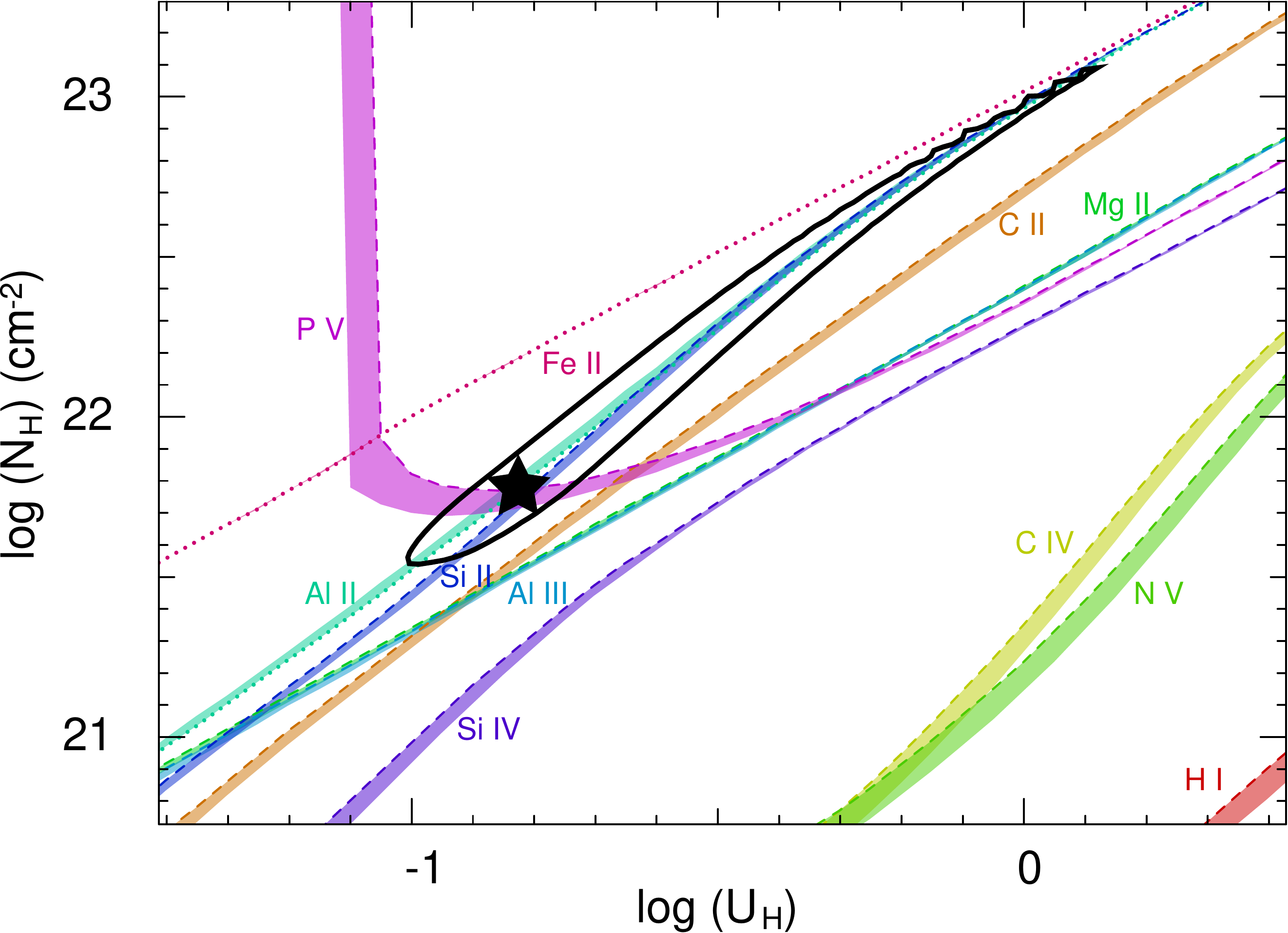}{0.33\textwidth}{(c)}}
    \caption{Plots of $\log{N_H}$ vs. $\log{U_H}$ for (a) S1, (b) S2, and (c) S3. The colored lines represent the $N_H$ and $U_H$ values allowed by the measured column densities of ions. Solid lines show measurements, dashed lines show lower limits, and dotted lines show upper limits. The colored bands attached to the lines represent the uncertainties in the column density measurements. The black stars in the plots show the solution for $N_H$ and $U_H$ found via $\chi^2$ minimization, and the black ellipses represent the $1\sigma$ range for the solutions. For this calculation, the HE0238 SED and solar metallicity are assumed.}
    \label{fig:nvuplot}
\end{figure*}

We use Cloudy to create a grid of simulated models that correspond to different $N_H$ and $U_H$ values, assuming solar metallicity, and the spectral energy distribution (SED) of quasar HE 0238-1904 (hereafter HE0238) \citep{2013MNRAS.436.3286A}. The $N_H$ and $U_H$ parameters determine the ionic column densities of each model, which we compare with the measured column densities shown in Table \ref{table:coldensity}. For S2, including the lower bound of the \ion{Fe}{2} column density in the analysis introduced an $N_H$ and $U_H$ solution that was contradictory to the constraints from the other ions. We suspect that this is because the Fe abundance of the system does not match solar metallicity ($Z_\odot$), requiring a metallicity of $\sim10Z_\odot$. This is in approximate agreement with the highest outflow metallicity found by \citet{2006ApJ...646..742G} ($Z\approx 5Z_\odot$). For this reason, we model our solution using the other ions but excluding \ion{Fe}{2}. The $\log{N_H}$ and $\log{U_H}$ values from this analysis are shown in Table \ref{table:energetics}, as well as in Figure \ref{fig:nvuplot}.\par
\subsection{Electron Number Density} \label{subsec:edensity}
The electron number density and, by extension, the distance of the outflow systems from the central source, can be found by determining the abundance ratios, measured via column densities, between excited and resonance states of low ionization species \citep{2009ApJ...706..525M}. We use the CHIANTI 9.0.1 Database \citep{1997A&AS..125..149D,Dere_2019} to model the relationship between the ratio of excited and resonance state ion abundances, and the electron number density, based on collisional excitation. We overlay this relation with the ratios based on the measured column densities, as shown in Figure \ref{fig:ratioplot}. For this object, we use the ratios $N(\text{\ion{Si}{2}*})/N(\text{\ion{Si}{2}})$, $N(\text{\ion{C}{2}*})/N(\text{\ion{C}{2}})$, and $N(\text{\ion{Fe}{2}*})/N(\text{\ion{Fe}{2}})$, where $N(ion)$ is the column density of a particular ion.\par
For S3, we have an upper limit given by the \ion{C}{2} ratio, and a measurement from the \ion{Si}{2} ratio which agree with one another. Our measurements of \ion{Fe}{2} are dominated by noise, and as such, are not included in the $n_e$ measurement. Taking the ratio of N(\ion{Si}{2}*)/N(\ion{Si}{2}), we find that $\log{n_e} = 3.3^{+0.8}_{-0.4}\:[\text{cm}^{-3}]$. S2 provides us a measurement from \ion{C}{2}, and upper limits from \ion{Si}{2} and \ion{Fe}{2}. From the N(\ion{C}{2}*)/N(\ion{C}{2}) ratio, we find $\log{n_e} = 0.25^{+0.2}_{-0.2}\:[\text{cm}^{-3}]$. S1 only gives us a lower limit from \ion{C}{2}, as the \ion{Si}{2}* is contaminated by an intervening line and cannot give us a reliable ratio between N(\ion{Si}{2}*) and N(\ion{Si}{2}). Thus, we get a lower limit for the electron number density, $\log{n_e} > 2.0_{-0.45}\:[\text{cm}^{-3}]$.
\section{Results} \label{sec:results}
\subsection{Distance and Kinetic Luminosity of the Outflows}\label{subsec:energetics}
In order to find the distance of the outflow systems, we use the definition for the ionization parameter
\begin{equation}
    \label{eq:UH}
    U_H \equiv \frac{Q_H}{4\pi R^2 n_H c}
\end{equation}
where $Q_H$ is the rate of ionizing photons, R is the distance of the outflow from the central source, and $n_H$ is the hydrogen number density, which is estimated as $n_e \approx 1.2 n_H$ for highly ionized plasma \citep{2006agna.book.....O}. Since we have a solution for $U_H$ from our photoionization analysis, as well as the $n_e$ for each outflow from the excited to resonance state ratios, we can find R after determining the value of $Q_H$. We determined $Q_H$ by first scaling the HE0238 SED to match the continuum flux at observed wavelength $\lambda=6500$ \AA\text{ }from the most recent SDSS observation ($F_\lambda=1.4^{+0.14}_{-0.14}\times10^{-16}$ erg s$^{-1}$ cm$^{-2}$ \AA$^{-1}$), and integrating over the scaled SED for energies above 1 Ryd, yielding $Q_H = 1.21^{+0.11}_{-0.11}\times10^{57} \text{ s}^{-1}$. The corresponding $L_{bol}=1.93^{+0.18}_{-0.18}\times 10^{47}$ erg s$^{-1}$ is larger than what would be expected from calculating the $\nu L_\nu$ at a specific wavelength via the method employed by \citet{2011MNRAS.410..860A}, as the HE0238 SED shows a large peak at the UV range \citep[$\lambda\approx1000$\text{ \AA,}][]{2013MNRAS.436.3286A}. Applying a bolometric correction appropriate to 1700 \AA\text{ }from \citet{2006ApJS..166..470R} brings the $\nu L_\nu$ reported by \citet{2011MNRAS.410..860A} to within 20\% of our calculated $L_{bol}$. The resulting outflow distances are shown in Table \ref{table:energetics}. Note that the distance of S2 (--1800 km s$^{-1}$, $R=67^{+55}_{-31} \text{ kpc}$) is at least an order of magnitude larger than that of S3 (--3500 km s$^{-1}$, $R=1.2^{+0.8}_{-0.9}$ kpc) or S1 (--1200 km s$^{-1}$, $R<5.4^{+7.3}$ kpc).\par
Once we have the distance of the outflow, we can find the mass flow rate \citep{2012ApJ...751..107B}
\begin{equation}
    \label{eq:Mdot}
    \dot{M} \simeq 4\pi\Omega R N_H \mu m_p v
\end{equation}
and the kinetic luminosity
\begin{equation}
    \label{eq:Ekdot}
    \dot{E}_k \simeq \frac{1}{2} \dot{M} v^2
\end{equation}
assuming a partially filled shell, where $\Omega$ is the global covering factor (fraction of the total solid angle of the quasar that the outflow covers), $\mu = 1.4$ is the mean atomic mass per proton, $m_p$ is the proton mass, and $v$ is outflow velocity. For the global covering factor, we assume $\Omega = 0.2$, the portion of quasars from which \ion{C}{4} BALs are found \citep{2003AJ....125.1784H}. As explained by \citet{2010ApJ...709..611D}, this is a reasonable assumption despite the relative rarity of quasars showing singly ionized absorption troughs such as \ion{Si}{2}, due to the likelihood that such quasars are regular BAL quasars seen from specific lines of sight. The resulting kinetic luminosity calculations yield $\log{\dot{E}_K} [\text{erg s}^{-1}] = 45.42^{+1.33}_{-0.64},45.82^{+0.37}_{-0.32}$ for S3 and S2 respectively, as well as an upper limit of $\log{\dot{E}_K} < 44.33^{+0.53}$ for S1. In addition, we calculate the momentum flux ($\dot{M}v$) of each outflow system (see Table \ref{table:energetics}) and compare it to the single-scattering limit of the quasar ($\frac{L_{bol}}{c}=6.44^{+0.61}_{-0.61}\times10^{36}\text{ erg cm}^{-1}$). The single-scattering limit assumes the scenario in which absorption of photon momentum drives acceleration \citep{1982ApJ...259..282A,1994ApJ...427..700A}. The momentum flux of S1 is smaller than the single-scattering limit, while those of S2 and S3 are above the limit. As S2 has a momentum flux an order of magnitude higher than the single-scattering limit, this implies the possibility of a multiple-scattering scenario \citep{1993ApJ...405..738L}.

\subsection{Changes in the High Velocity BAL Trough (S4)}\label{subsec:velocityshift}
Following up on the results reported by \citet{2007ApJ...665..174H}, we examine the velocity shift of the \ion{C}{4} BAL of S4. Using two Gaussian profiles, one broad and shallow, and the other narrow and deep, we modeled the absorption in each of the five epochs, as shown in Figure \ref{fig:bigBALvelocities}. We can see that the centroid velocity of the narrow Gaussian monotonically grows, while the equivalent width becomes smaller from epoch to epoch. Detailed information on the centroid velocities and equivalent widths per epoch can be seen in Table \ref{table:epoch_velocity}.\par
Assuming acceleration along the line of sight, based on the centroid velocities of the narrow Gaussian, the average acceleration between the observations in September 2001 and January 2017 would be $a = -0.25\pm{0.13} \text{ cm s}^{-2}$ in the quasar's rest frame, which agrees within error with the acceleration $a = -0.154 \pm{0.025}\text{ cm s}^{-2}$ between September 2001 and September 2005 found by \citet{2007ApJ...665..174H}. Due to the shrinking of the trough, we must take into consideration effects other than line of sight acceleration, such as changes in photoionization, as discussed by \citet{2020ApJS..247...40X}.\par
\begin{figure}
    \centering
    \plotone{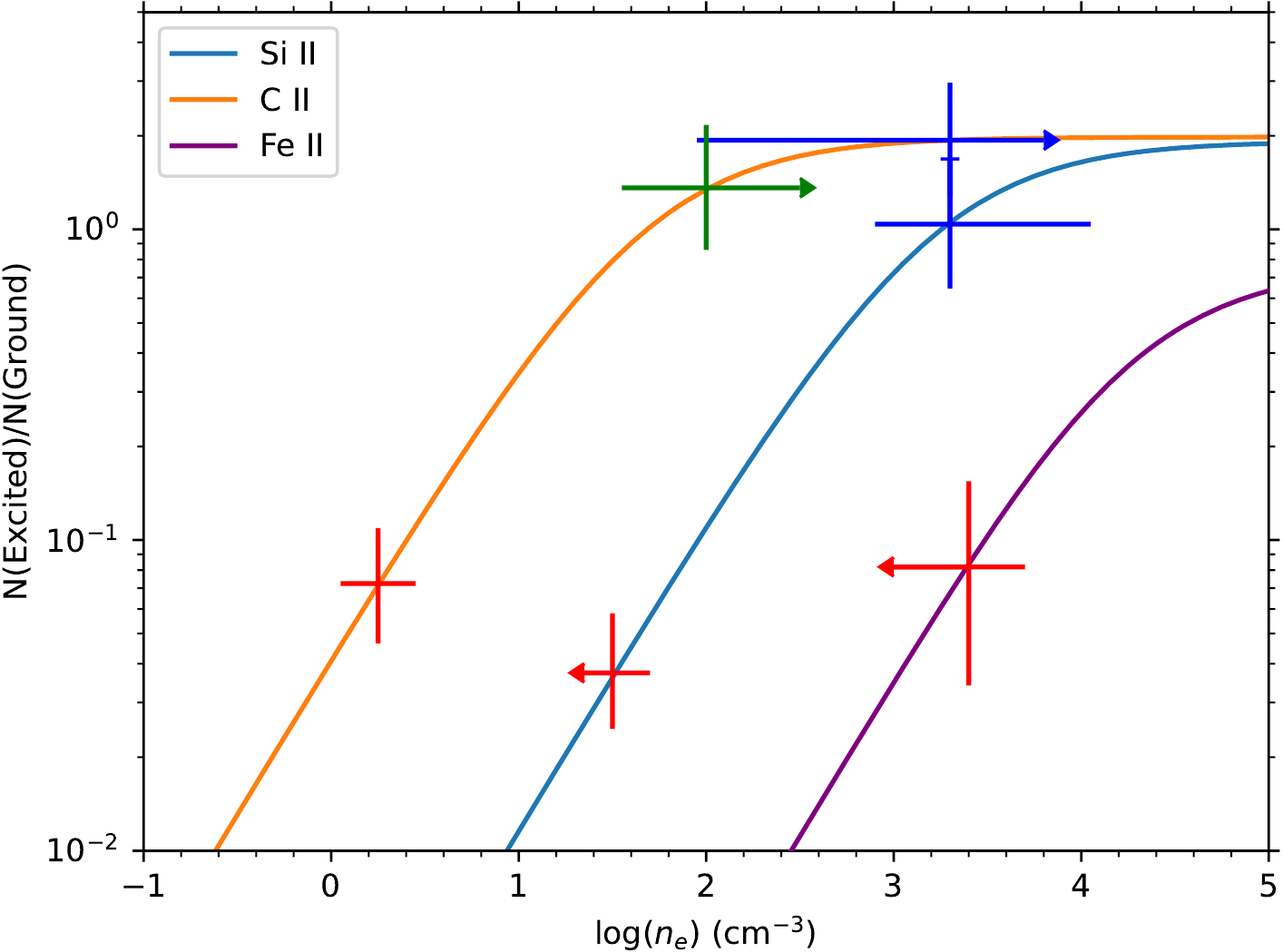}
    \caption{Ratio between excited and resonance state abundances of \ion{Si}{2}, \ion{C}{2}, and \ion{Fe}{2} vs. $\log{n_e}$. The curves marked \ion{Si}{2}, \ion{C}{2}, and \ion{Fe}{2} are the theoretical ratios modeled with CHIANTI, assuming a temperature of 10,000 K. The crosses on the curves show the ranges of the \ion{C}{2}, \ion{Si}{2}, and \ion{Fe}{2} column density ratios, based on the measured AOD column densities. The green, red, and blue correspond to systems S1, S2, and S3 respectively. Arrows indicate either upper or lower limits in $\log{n_e}$ depending on the direction of the arrow. The upper limit of the N(\ion{Si}{2}*)/N(\ion{Si}{2}) ratio for S3 is marked with a tick, as it overlaps with the error bars of the \ion{C}{2} ratio of the same system.}
    \label{fig:ratioplot}
\end{figure}
\begin{deluxetable*}{lccc}
\tablecaption{Physical Properties J0242+0049 Outflow Systems}
\tablenum{3}
\tablehead{\colhead{Outflow System} & S1 = $-1200 \text{ km s}^{-1}$&S2 = $-1800 \text{ km s}^{-1}$&S3 = $-3500 \text{ km s}^{-1}$} 

\startdata
\vspace{-0.2cm}$log(N_{\text{H}})$&&&\\\vspace{-0.2cm}
&$21.41_{-0.70}^{+0.38}$&$21.27^{+0.64}_{-0.58}$&$21.78^{+1.30}_{-0.24}$\\
$[\text{cm}^{-2}]$&&&\\
\hline
\vspace{-0.2cm}$log(U_{\text{H}})$&&&\\\vspace{-0.2cm} &$-0.86_{-0.59}^{+0.33}$&$-1.30^{+0.49}_{-0.48}$&$-0.83^{+0.95}_{-0.18}$\\
$[\text{dex}]$&&&\\
\hline
\vspace{-0.2cm}$log(n_{\text{e}})$&&&\\\vspace{-0.2cm} &$>2.00_{-0.45}$&$0.25^{+0.20}_{-0.20}$&$3.30^{+0.75}_{-0.40}$\\
$[\text{cm}^{-3}]$&&&\\
\hline
\vspace{-0.2cm}$\text{Distance}$&&&\\\vspace{-0.2cm}&$<5.4^{+7.3}$&$67^{+55}_{-31}$&$1.2^{+0.8}_{-0.9}$\\
$[\text{kpc}]$&&&\\
\hline
\vspace{-0.2cm}$\dot M$&&&\\\vspace{-0.2cm}&$<480^{+300}$&$6500^{+8900}_{-3400}$&$700^{+2900}_{-30}$\\
$[M_{\odot} \text{yr}^{-1}]$&&&\\
\hline
\vspace{-0.2cm}$\dot M v$&&&\\\vspace{-0.2cm}&$<3.6^{+2.3}$&$74^{+100}_{-39}$&$16^{+60}_{-0.7}$\\
$[10^{36}\text{ erg cm}^{-1}]$&&&\\
\hline
\vspace{-0.2cm}$log({\dot E}_K)$&&&\\\vspace{-0.2cm}&$<44.33^{+0.21}$&$45.82^{+0.37}_{-0.32}$&$45.43^{+0.7}_{-0.02}$\\
$[\text{erg s}^{-1}]$&&&\\
\hline
\vspace{-0.2cm}${\dot E}_K/L_{edd}$&&&\\\vspace{-0.2cm}&$<0.18^{+0.16}$&$5.5^{+8.8}_{-3.1}$&$2.3^{+9.9}_{-0.8}$\\
$[\text{\%}]$&&&\\
\enddata
\label{table:energetics}
\tablecomments{Temperature of 10,000K assumed.}
\end{deluxetable*}
\begin{deluxetable*}{lcccccccc}
\tablenum{4}
\tablecaption{Velocities of \ion{C}{4} BAL at Each Epoch\label{table:epoch_velocity}}
\tabletypesize{\scriptsize}
\tablehead{\colhead{\vspace{-0.3cm}MJD}&\colhead{Date}&\colhead{$\Delta t_{Rest}$}&\colhead{$v_{n}$}&\colhead{$\Delta v_{n}$}&\colhead{$EW_n$}&\colhead{$v_{w}$}&\colhead{$\Delta v_{w}$}&\colhead{$EW_w$}\\
\colhead{}&\colhead{}&\colhead{(days)}&\colhead{($\text{km s}^{-1}$)}&\colhead{($\text{km s}^{-1}$)}&\colhead{($\text{km s}^{-1}$)}&\colhead{($\text{km s}^{-1}$)}&\colhead{($\text{km s}^{-1}$)}&\colhead{($\text{km s}^{-1}$)}}
\startdata
\text{52177}&Sep. 25, 2001&$0$&$-17,460\pm{50}$&$0$&$1190$&$-19,000$&$0$&$1010$\\
\text{52199}&Oct. 17, 2001&13.0&$-17,600\pm{40}$&$-140\pm{70}$&$1320$&$-19,000$&$0$&$2140$\\
\text{53619}&Sep. 5, 2005&838.6&$-17,720\pm{4}$&$-260\pm{50}$&$920$&$-20,000$&$-1000$&$20$\\
\text{55455}&Sep. 16, 2010&1922.8&$-17,870\pm{40}$&$-400\pm{60}$&$460$&$-19,490$&$-490$&$860$\\
\text{57758}&Jan. 5, 2017&3282.7&$-18,180\pm{380}$&$-720\pm{380}$&$80$&$-19,000$&$0$&$260$\\
\enddata
\tablecomments{Table of the centroid velocity and equivalent width of the \ion{C}{4} BAL for each epoch. $\Delta t_{Rest}$ is the elapsed time in the quasar's rest frame since the 52177 epoch. $v_n$ and $v_w$ are the centroid velocities of the narrow and wide best fit Gaussians in the quasar's rest frame, while $\Delta v_n$ and $\Delta v_w$ are the velocity shifts compared to that of the 52177 epoch. The equivalent widths ($EW_n, EW_w$) have been calculated by integrating over the Gaussians in velocity space. The parameters for the wider Gaussians are more affected by the continuum models for each epoch. Note that the uncertainty in the centroid velocity of the MJD=53619 epoch is significantly smaller than those of the other epochs, due to the higher S/N ratio and resolution of the data.}
\end{deluxetable*}
\begin{figure}
    \centering
    \gridline{\fig{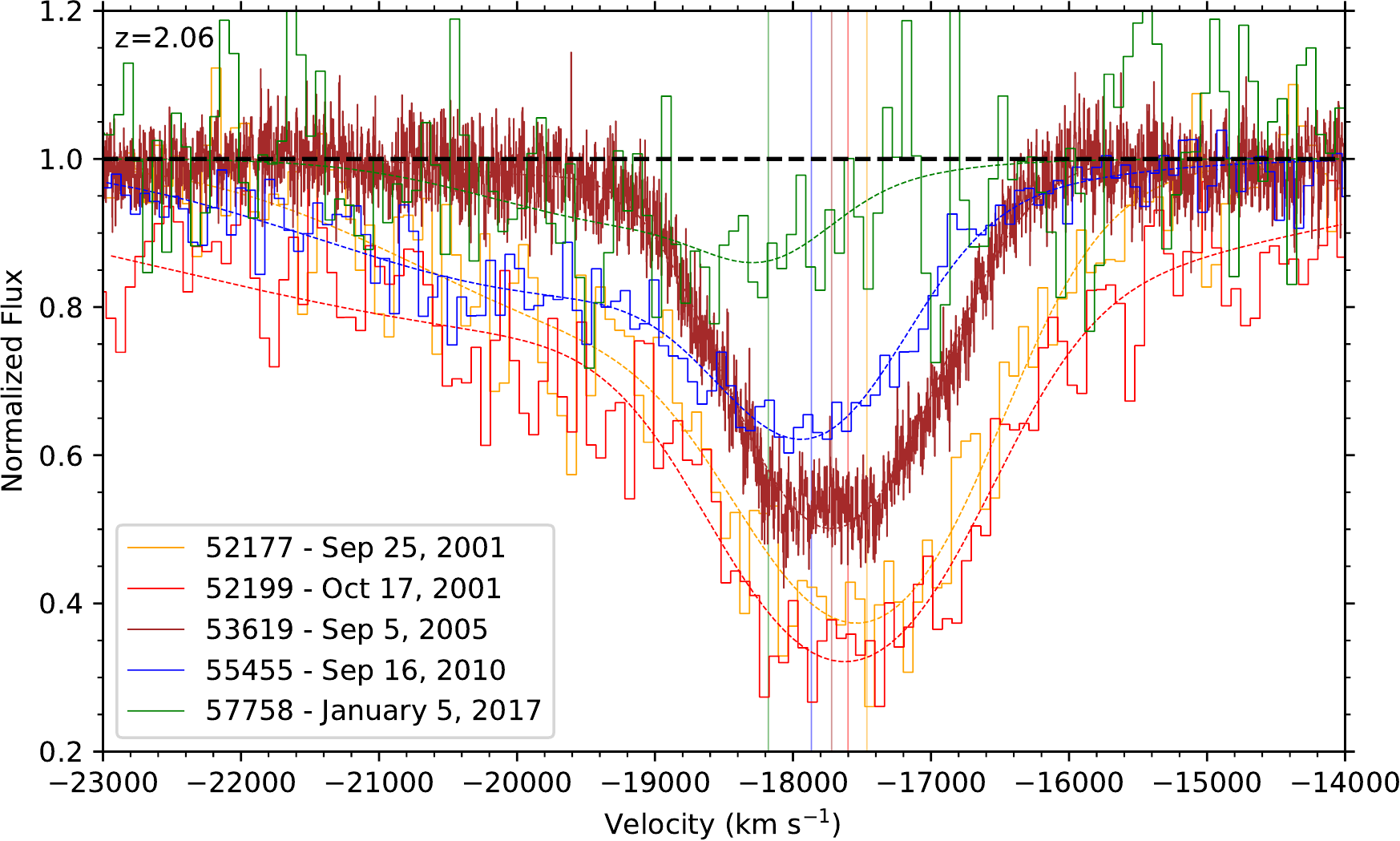}{0.4\textwidth}{(a)}}
    \gridline{\fig{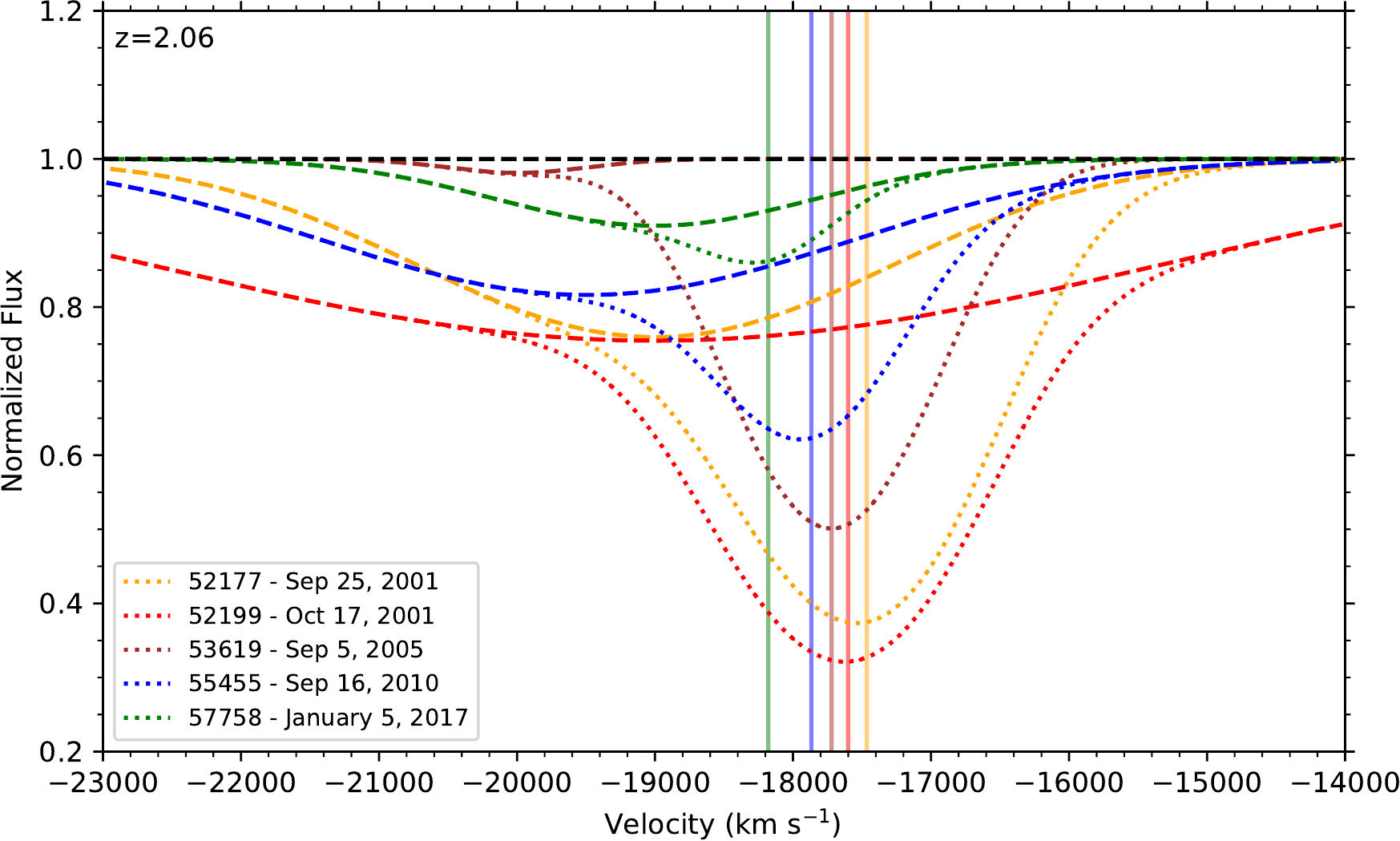}{0.4\textwidth}{(b)}}
    \caption{Normalized flux vs. velocity of the S4 \ion{C}{4} BAL at different epochs. The trough has been modeled by employing a best fit of a profile of two Gaussians, one wide and one narrow. (a) shows Gaussian models of the troughs over the data, while (b) shows the Gaussian models independently. The colored vertical lines mark the centroid velocities at each epoch. Note that the centroid velocity increases through each epoch, while the EW decreases. }
    \label{fig:bigBALvelocities}
\end{figure}
\section{Discussion} \label{sec:discussion}
\subsection{AGN Feedback Contribution of Outflows}
As previously mentioned in the introduction, the kinetic luminosity ($\dot{E}_k$) of the outflow systems must be at least $\sim0.5\%$ \citep{2010MNRAS.401....7H} or $\sim5\%$ \citep{2004ApJ...608...62S} of the source quasar's Eddington luminosity ($L_{Edd}$) to contribute to AGN feedback. In order to find this ratio, we must first find the Eddington luminosity. We compute the mass of the black hole using the \ion{Mg}{2}-based mass equation in \citet{2019ApJ...875...50B}, with the FWHM of the \ion{Mg}{2} emission feature in the SDSS spectrum. To account for the \ion{Fe}{2} emission throughout the spectrum, we use the \ion{Fe}{2} template by \citet{2006ApJ...650...57T}, and run a best fit algorithm to match the features in the spectrum, as done by \citet{2018ApJ...859..138W}. This yields a black hole mass of $M_{BH} = 9.7^{+4.9}_{-3.4}\times10^8 M_\odot$, corresponding to an Eddington luminosity of $L_{Edd} = 1.2^{+0.6}_{-0.4}\times10^{47} \text{erg s}^{-1}$. We expect the \ion{Fe}{2} emission's effect on the absorption to be small, as the fitted emission template from \citet{2006ApJ...650...57T} is $<20\%$ of the continuum level of the SDSS spectrum of MJD=57758, leaving us with column densities that agree with our measured values within error.\par
Taking the ratio between the kinetic luminosity of each outflow system and the Eddington luminosity of the quasar, we find that S2 and S3 are well above the 0.5\% threshold from \citet{2010MNRAS.401....7H}, and S2 is above the 5\% threshold by \citet{2004ApJ...608...62S}, while S1's kinetic luminosity is below 0.18 \% of the Eddington luminosity, as seen in Table \ref{table:energetics}. We can thus conclude that S2 and S3 are energetic enough to contribute to AGN feedback.\par
Unlike in objects analyzed in other papers \citep[e.g.][]{2020ApJS..247...39M,2020ApJS..247...38X}, we do not have lines from the very high ionization phase. Thus, while there may be a very high ionization phase, we cannot tell from the information we have.
\subsection{Time Variability of Troughs}
\begin{figure}
    \centering
    \plotone{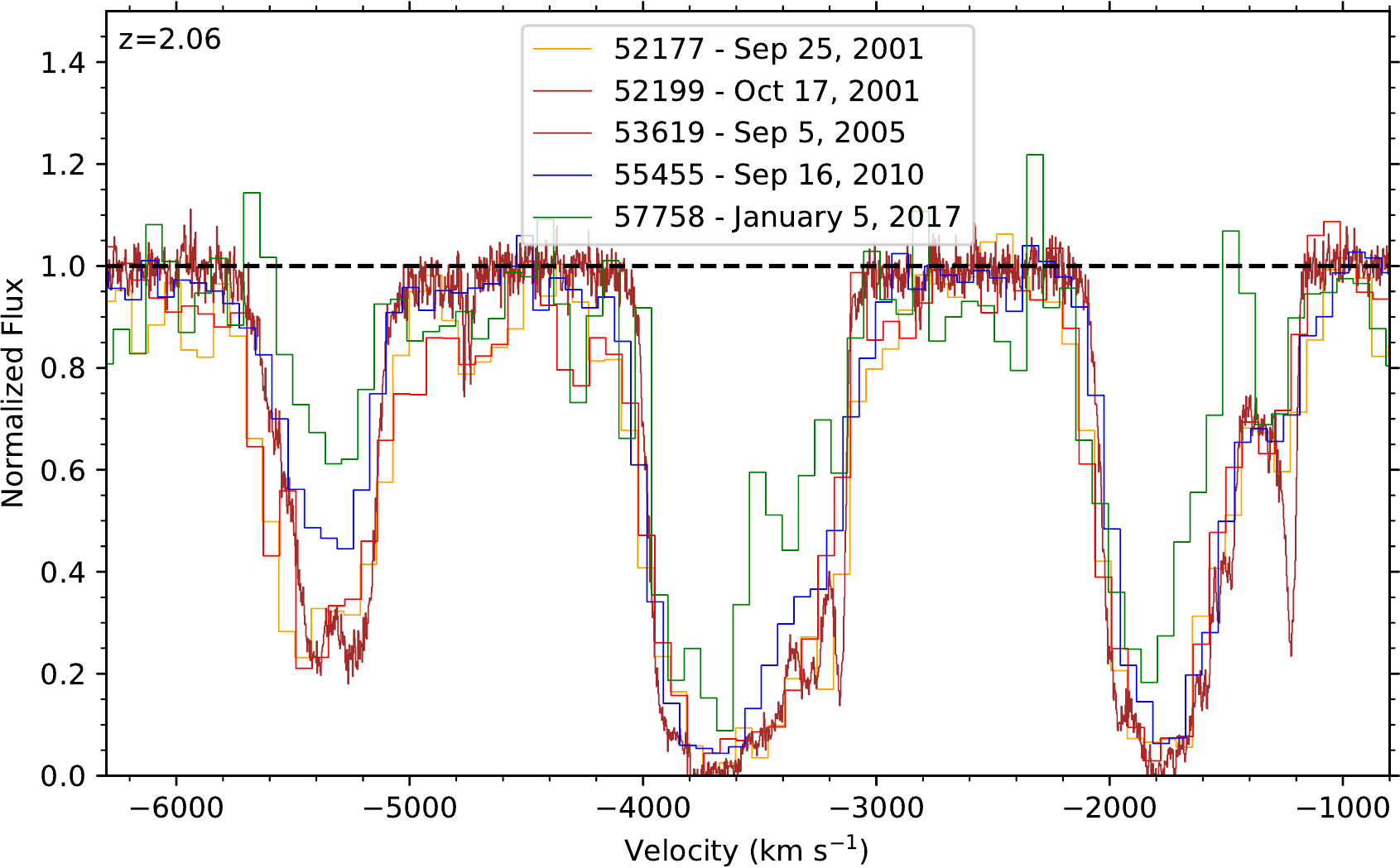}
    \caption{Normalized flux vs. velocity of \ion{Si}{4} troughs at different epochs. Note that the depth of the troughs becomes shallower over time.}
    \label{fig:SiIV_multiepoch}
\end{figure}

Following the examination of the S4 \ion{C}{4} BAL at different epochs, we looked to systems S1, S2, and S3 for time variability. As shown in Figure \ref{fig:SiIV_multiepoch}, the \ion{Si}{4} trough depth becomes increasingly shallower over time, which may be explained by the same ionization effects that affect the S4 \ion{C}{4} BAL shown in Figure \ref{fig:bigBALvelocities}, discussed by \citet{2020ApJS..247...40X}. As the ionization parameter $U_H$ changes, ions of particular ionization states become more or less abundant over time. Since the \ion{C}{4} of S4, along with \ion{Si}{4} of S1, S2, and S3, decrease monotonically, this supports the assertion that the changes in the troughs are due to changes in the ionization parameter. Further observation and analysis will be required to confirm these effects.
\subsection{SED and Metalliticy Dependency, and Attenuation of the SED}
An alternative to using the SED of HE 0238-1904 would be to use the theoretical SED as defined by \citet{1987ApJ...323..456M}, which is based on the He II line. The HE 0238-1904 SED is based on observation of a high quality spectrum which stretches into the far UV range, better representing a quasar spectrum \citep{2013MNRAS.436.3286A}.
Just like in other objects\citep[e.g.][]{2018ApJ...858...39X,2020ApJS..247...39M}, higher metallicity drops the values of the energetics parameters. For instance, raising the metallicity to 4 times solar metallicity, using abundance ratios from \citet{2008A&A...478..335B}, changes the photoionization solution of S2 to $\log{U_H}=-1.5^{+0.3}_{-0.3}$, and $\log{N_H}=20.5^{+0.4}_{-0.4} [\text{cm}^{-2}]$, lowering the mass flow rate and kinetic luminosity to $\dot{M}=1300^{+600}_{-400} M_\odot \text{yr}^{-1}$ and $\log{\dot{E}_K}=45.13^{+0.17}_{-0.18} [\text{erg s}^{-1}]$ respectively. Using the SED by \citet{1987ApJ...323..456M} with solar metallicity changes the solution to $\log{U_H}=-1.5^{+0.4}_{-0.4}, \log{N_H}=21.2^{+0.5}_{-0.5}$, which is in agreement with the values in Table \ref{table:energetics} within error.\par
It is possible that the SED seen by one outflow system can be attenuated by another, resulting in a smaller $Q_H$, and by extension, a smaller distance $R$. In particular, as S2 is further out than the other mini-BAL system S3, it is likely that the SED seen by S2 is obscured by S3 \citep[e.g.,][]{2010ApJ...713...25B,2017ApJ...838...88S,2018ApJ...865...90M,2020ApJS..247...41M}. We used the method described by \citet{2018ApJ...865...90M} to test the effects of attenuation by S3. We used Cloudy to model the attenuated SED by S3 by inputting the relevant $N_H$ and $U_H$ values of S3 shown in Table \ref{table:energetics}. We then use that attenuated SED to find the resulting $Q_H$ and $R$ of S2. The reduced values for the parameters are $Q_H=4.9^{+0.5}_{-0.5}\times10^{56}\text{ s}^{-1}$ and $R=43^{+35}_{-20}$ kpc, which is a $\sim30\%$ decrease in the distance of S2. We choose S3 as the attenuation source, as its stronger features compared to S1 suggest that the attenuation effect from S3 would be larger than that of S1. We are unable to calculate the attenuation by S4, as we cannot obtain $N_H$ or $U_H$ from its singular \ion{C}{4} absorption trough.

\subsection{Comparison With Other Outflows}
There have been several prior studies of quasar outflow acceleration, including that of the acceleration of the outflow of quasar SDSS J1042+1646 conducted by \citet{2020ApJS..247...40X}, based on the acceleration seen in the \ion{Ne}{8} $\lambda\lambda 770, 780$. The bolometric luminosity of SDSS J1042+1646 is estimated at $\sim1.5\times10^{47}\text{ erg s}^{-1}$, which is comparable to that of J0242+0049 ($1.9\times10^{47}\text{ erg s}^{-1}$). The average acceleration of S4 that we have found ($a\approx-0.25 \text{ cm s}^{-2}$) is roughly an order of magnitude smaller than that by \citet{2020ApJS..247...40X} ($a=-1.52 \text{ cm s}^{-2}$), which suggests that if S4 is truly accelerating, the acceleration of quasar outflows can cover a wide range.\par
To give context to the study of outflow S2, we review a few outflows with similarly large R and/or $\dot{E}_K$. Analysis of a molecular outflow of quasar SDSS J1148+5251 at a distance $R\sim 15\text{ kpc}$ conducted by \citet{2012MNRAS.425L..66M} revealed a lower limit to the mass flow rate of $\dot{M}>3500 M_{\odot}\text{ yr}^{-1}$ as well as one for the kinetic luminosity $\dot{E}_K > 1.9\times10^{45}\text{ erg s}^{-1}$.\par
\citet{2013MNRAS.436.2576L} analyzed the ionized gas around 11 radio-quiet quasars via the [\ion{O}{3}] $\lambda5007\text{\AA}$ emission. These outflows were found between $\sim10-20\text{ kpc}$ from the central source, had velocities of up to --1000 km s$^{-1}$, and had an estimated $n_e\sim1.2\text{ cm}^{-3}$. The outflows had an estimated range of $\dot{E}_K$ from $4\times10^{44}$ to $3\times10^{45}$ erg s$^{-1}$, and $\dot{M}$ from $2\times10^3$ to $2\times10^4 M_{\odot}\text{ yr}^{-1}$. These numbers are within a factor of few of the values we find for S1, S2, and S3 (see Table \ref{table:energetics}), which suggests we may find similar outflows in absorption.\par
In their analysis of SDSS J1051+1247, \citet{2020ApJS..247...39M} found an outflow system with $\dot{E}_K =3\times10^{45}\text{ erg s}^{-1}$. The mass flow rate ($\dot{M}=6500M_{\odot}\text{ yr}^{-1}$) and kinetic luminosity ($\dot{E}_K = 6.6\times10^{45}\text{ erg s}^{-1}$) of S2 align with these values, and those of the objects mentioned above, within margin of error. \citet{2020ApJS..247...38X} claim the most energetic quasar outflow measurement to date from quasar SDSS J1042+1646 ($\dot{E}_K = 5\times10^{46}\text{ erg s}^{-1}$), and this claim remains uncontested.\par
While the distance of S2 from the quasar is unprecedentedly large, there exists a theoretical model that may be supported by this observation. \citet{2012MNRAS.420.1347F} provide an argument that FeLoBALs, absorption systems with signs of \ion{Fe}{2}, may be formed in situ at distances of several kpc. They clarify that while their model focuses on the formation of FeLoBALs at large distances, other classes of outflows may form as described by it.
\section{Summary and Conclusion} \label{sec:conclusion}
This paper has presented the analysis of three absorption systems of quasar SDSS J0242+0049, dubbed S1, S2, and S3, from VLT/UVES observational data, as well as the velocity shift of the S4 \ion{C}{4} BAL across five different epochs. From the absorption troughs we identified, we measured the column densities of 11 ions in each system as shown in Table \ref{table:coldensity}. Through photoionization analysis using the measured column densities, we found the best fit solutions to $U_H$ and $N_H$ for each system.\par
The abundance ratios between the excited and resonance states of ions \ion{Si}{2} and \ion{C}{2} were used to find the electron number density $n_e$ of the three systems S1, S2, and S3, as shown in Figure \ref{fig:ratioplot}. Equations \ref{eq:UH}, \ref{eq:Mdot}, and \ref{eq:Ekdot} used to find the distance from the central source, the mass flow rate, and the kinetic luminosity of each system respectively. The ratios between the kinetic luminosities and the quasar's Eddington luminosity were found in order to evaluate their AGN feedback contribution, the results of which can be seen in Table \ref{table:energetics}. From this analysis, we have found that S2 and S3 have sufficient kinetic luminosity for AGN feedback contribution. Most notable in this result is the distance of S2 $R=67$ kpc, further than the absorption system of 3C 191 found at R= 28 kpc by \citet{2001ApJ...550..142H}, making this the furthest reported distance of a mini-BAL absorption outflow from its central source.\par
Following the analysis of the three systems, we examined the change in velocity and equivalent width of the S4 \ion{C}{4} BAL, as shown in Figure \ref{fig:bigBALvelocities}, based on the UVES spectrum, as well as different SDSS observations. As seen in Table \ref{table:epoch_velocity}, there has been a monotonic increase in the line of sight velocity, as well as a decrease in equivalent width, with the trough being a factor of six weaker at the epoch of January 2017 compared to that of September 2001.\par
Through further observation and analysis, we expect to shed more light on the time variability of the S4 \ion{C}{4} BAL, as well as that of systems S1, S2, and S3.\par
\begin{acknowledgments}
NA and DB acknowledge support from NSF grant AST 2106249, as well as NASA STScI grants GO 14777, 14242, 14054, 14176, and AR-15786. DB acknowledges support from the Virginia Space Grant Consortium Graduate Research Fellowship Program.
\end{acknowledgments}

\bibliography{j0242}{}
\bibliographystyle{aasjournal}



\end{document}